\documentclass[aps,prb,amsmath,amssymb,floatfix,citeautoscript,showpacs,reprint]{revtex4-1}

\usepackage[colorlinks=true,citecolor=blue,urlcolor=blue,linkcolor=blue]{hyperref}
\usepackage{amsmath,amssymb,amsthm}
\usepackage{bbm}
\usepackage[pdftex]{color} 
\usepackage{graphicx}
\usepackage{bm}
\usepackage{multirow}         
\usepackage{tikz}
\usetikzlibrary{calc}

\newcommand{\bs}[1]{{\boldsymbol{#1}}} 
\newcommand{\E}{\mathbbm{1}}
\renewcommand{\d}{\mathrm{d}}
\renewcommand{\i}{\mathrm{i}}
\newcommand{\e}{\mathrm{e}}

\newcommand{\zz}{\mathbbm{Z}_2}
\newcommand{\bk}{\bs{k}}
\newcommand{\hc}{\text{H.c.}}

\newcommand{\bra}[1]{\left\langle #1\right|}
\newcommand{\ket}[1]{\left| #1\right\rangle}
\renewcommand{\mathbb}{\mathbbm}
\newcommand{\Nnn}[1]{\langle\!\langle #1 \rangle\!\rangle}
\newcommand{\Nnnn}[1]{\langle\!\langle\!\langle #1 \rangle\!\rangle\!\rangle}
\newcommand{\dd}[2]{\frac{\partial #1}{\partial #2}}

\newcommand{\fig}[1]{Fig.~\ref{#1}}
\newcommand{\app}[1]{Appendix~\ref{#1}}
\newcommand{\Sec}[1]{Sec.~\ref{#1}}
\newcommand{\eq}[1]{Eq.~\eqref{#1}}
\newcommand{\tab}[1]{Table~\ref{#1}}
\newcommand{\Ref}[1]{Ref.~\onlinecite{#1}}

\renewcommand{\tt}{t}
\newcommand{\ttt}{t'}
\newcommand{\tttt}{t''}
\newcommand{\td}{t_{d}}
\newcommand{\tdd}{t_{d}'}
\newcommand{\tddd}{t_{d}''}
\newcommand{\tf}{t_{f}}
\newcommand{\tff}{t_{f}'}
\newcommand{\tfff}{t_{f}''}
\newcommand{\ef}{\epsilon_{f}}
\newcommand{\nf}{n_{f}}
\newcommand{\nd}{D}

\newcommand{\abb}{\textsc}
\newcommand{\trim}{\abb{trim}}
\newcommand{\hsp}{\abb{hsp}}
\newcommand{\hsl}{\abb{hsl}}
\newcommand{\bz}{\abb{bz}}
\newcommand{\sbz}{s\abb{bz}}
\newcommand{\nn}{\abb{nn}}
\newcommand{\nnn}{\abb{nnn}}
\newcommand{\nnnn}{\abb{nnnn}}
\newcommand{\wwti}{\abb{wti}}
\newcommand{\ssti}{\abb{sti}}
\newcommand{\bi}{\abb{bi}}
\newcommand{\ttci}{\abb{tci}}

\newcommand{\smb}{$\rm SmB_6$}
\newcommand{\pt}[1]{\mathrm{#1}}
\newcommand{\ptt}[1]{$\pt{#1}$}
\newcommand{\wti}[1]{\wwti(\ptt{#1})}
\newcommand{\sti}[1]{\ssti(\ptt{#1})}
\newcommand{\tci}[1]{\ttci(\ptt{#1})}
\newcommand{\wtii}[1]{\wti{\overline{#1}}}
\newcommand{\stii}[1]{\sti{\overline{#1}}}

\newcommand{\figlabel}[3]{%
\begin{tikzpicture}[inner sep=0pt]
\node [anchor=north west] at (0,0) {#1};
\node [anchor=north west] at (#3) {(#2)};
\end{tikzpicture}}

\begin{document}

\title{Topological invariants, surface states, and interaction-driven phase transitions in correlated Kondo insulators with cubic symmetry}

\author{Markus Legner} 
\author{Andreas R\"uegg} 
\author{Manfred Sigrist} 
\affiliation{
Institut f\"ur Theoretische Physik, ETH Z\"urich,
8093 Z\"urich, Switzerland}  

\pacs{71.27.+a, 73.20.At, 71.10.Fd, 03.65.Vf}

\date{\today}

\begin{abstract}
We construct a lattice model for a cubic Kondo insulator consisting of one spin-degenerate $d$  and $f$ orbital at each lattice site. The odd-parity hybridization between the two orbitals permits us to obtain various trivial and topological insulating phases, which we classify in the presence of cubic symmetry. In particular, depending on the choice of our model parameters, we find a strong topological insulator phase with a band inversion at the \ptt{X} point, modeling the situation potentially realized in \smb, and a topological crystalline insulator phase with trivial $\zz$ indices but nonvanishing mirror Chern numbers.
Using the Kotliar-Ruckenstein slave-boson scheme, we further demonstrate how increasing interactions among $f$ electrons can lead to topological phase transitions. Remarkably, for fixed band parameters, the $f$-orbital occupation number at the topological transitions is essentially independent of the interaction strength, thus yielding a robust criterion to discriminate between different phases.
\end{abstract}

\maketitle

\section{Introduction}
Identifying novel material systems with topological band properties is of central relevance should the full potential of these phases be utilized in experiments and future applications.\cite{hasan_colloquium_2010,qi_topological_2011} Thereby, considerable effort has been devoted to detect and theoretically describe ``correlated topological insulators", i.e., systems where topological band features coexist with strong electron-electron interactions (see \Ref{hohenadler_correlation_2013} and references therein). The peculiar interplay between topology and interactions is interesting from at least two perspectives. On the one hand, novel interacting bulk phases with or without symmetry-protected boundary modes may be stabilized by the electron-electron interactions. On the other hand, electron-electron interactions may trigger phase transitions at the surface of a topological insulator, thereby exposing the nontrivial features of the bulk.

In this context, topological Kondo insulators\cite{ghaemi_higher_2007,dzero_topological_2010,dzero_theory_2012} emerged as promising systems to study interaction effects in topological insulators. Much of the current interest in this class of materials has been triggered by several recent experiments on samarium hexaboride (\smb), which consistently reported the coexistence of metallic surface states with an insulating bulk.\cite{miyazaki_momentum_2012,wolgast_low_2013,kim_robust_2013,zhang_hybridization_2013,xu_surface_2013,neupane_surface_2013,jiang_observation_2013,li_quantum_2013,yee_imaging_2013,frantzeskakis_kondo_2013,min_importance_2013} This behavior is in line with theoretical studies \cite{takimoto_smb6_2011,tran_phase_2012,lu_correlated_2013,alexandrov_cubic_2013,dzero_new_2013,ye_tci_2013,deng_plutonium_2013,weng_correlated_2013} predicting that \smb\ and other mixed-valence materials realize topological insulator phases with symmetry-protected metallic surfaces. But despite the agreement between theory and experiment, open questions, such as the origin of the observed small effective mass of surface electrons,\cite{li_quantum_2013,frantzeskakis_kondo_2013} remain. It was also suggested that other (nontopological) contributions to the surface transport are relevant,\cite{zhu_polarity_2013} rendering the field of topological Kondo insulators an active research area.

From the theoretical point of view, the crucial ingredient to realize a topological Kondo insulator (with inversion symmetry) is that the hybridization matrix $\Phi({\bs k})$ between the localized and itinerant electrons is an {\em odd} function of ${\bs k}$.\cite{dzero_topological_2010} This means that the hybridization is necessarily {\em non-onsite} and implies that localized and itinerant degrees of freedom have opposite parity. Consequently, if the system has inverted bands at an odd number of time-reversal-invariant momenta (\trim), a topological insulator is realized.\cite{fu_topological_2007} We note that an odd-parity hybridization is generic for many mixed-valance compounds as the localized degrees of freedom are typically derived from the atomic $f$ orbitals, which are odd under parity, and the itinerant degrees of freedom from the atomic $d$ orbitals with even parity.

Motivated by these recent activities, we construct and analyze in this paper a cubic lattice model for a Kondo insulator. It consists of two spin-degenerate orbitals which couple via an odd-parity hybridization, modeling, e.g., the localized $f$  and itinerant $d$ electrons of Sm in \smb.\cite{takimoto_smb6_2011,dzero_theory_2012} As a result of the cubic symmetry, we find that eight different gapped band insulators (with their respective charge-conjugate partners) are possible at half filling, realizing various trivial and topological phases. These different phases are all distinguished by their inversion eigenvalues at the eight \trim\ and also differ in the nature of their surface states (if present). We show that a complementary classification of these phases is possible using two mirror Chern numbers\cite{teo_surface_2008} (which require the existence of mirror planes) instead of the inversion eigenvalues. These mirror Chern numbers also uniquely determine (a) the strong $\zz$ invariant $\nu_0$ (protected by time-reversal symmetry and charge conservation) and (b) the three weak $\zz$ invariants ($\nu_1,\nu_2,\nu_3)$ (which require additional translation symmetry).

We consider the effect of electron-electron interactions among the $f$ electrons within the quasiparticle approximation assuming a ``local Fermi liquid".\cite{Read:1983b,Coleman:1984,Rice:1985b,Yamada:1986,kotliar_new_1986} The quasiparticle excitations of the interacting system are then accurately described by a renormalized quadratic Hamiltonian which can be classified topologically. In contrast to previous studies where the $U\to\infty$ approximation was applied,\cite{takimoto_smb6_2011,dzero_theory_2012,tran_phase_2012} we use the Kotliar-Ruckenstein slave-boson scheme at finite $U$~\cite{kotliar_new_1986} to compute the renormalization factors in a self-consistent manner and demonstrate that interactions can drive topological phase transitions.\cite{wang_interaction_2012,budich_fluctuation_2012,budich_fluctuation_2013,werner_interaction_2013,werner_temperature_2013}
Remarkably, for fixed band parameters, the $f$-orbital occupation number at the topological transitions is essentially independent of the interaction strength, thus yielding a robust criterion to discriminate between different phases.

The remainder of this paper is organized as follows. In \Sec{sec:model}, we introduce the model and discuss its noninteracting band structure as well as the renormalization effects in the presence of electron-electron interactions. In \Sec{sec:top}, we show how different topological phases emerge from band inversions at high-symmetry points, characterize them by different symmetry-protected topological invariants, and show how gapless surface modes are protected by the different symmetries. In \Sec{sec:interactions}, we discuss interactions among $f$-electrons and show how they can lead to topological phase transitions.

\section{Model}\label{sec:model}
\subsection{Cubic topological Kondo insulator}
Let us start by defining a minimal model for a (topological) Kondo insulator on a simple cubic lattice with one spin-degenerate orbital per lattice site each for $d$  and $f$-electrons.\cite{dzero_theory_2012} The general Hamiltonian has the form of the periodic Anderson model and is given by
\begin{equation}
H=H_0+H_{\rm hyb}+H_{\rm int}\,,\label{eq:model_general}
\end{equation}
where $H_0$, $H_{\rm hyb}$, and $H_{\rm int}$ describe the tight-binding energy of $d$  and $f$-electrons, the hybridization between $d$  and $f$-electrons, and the interactions, respectively. We include up to third-neighbors hopping in $H_0$ and assume an imaginary and spin-dependent hybridization between nearest-neighboring $d$  and $f$-electrons. Furthermore, we assume that the $f$ electrons locally interact via a Hubbard-$U$ repulsion while the $d$ electrons are noninteracting. Therefore, we can write the individual parts of the total Hamiltonian~\eqref{eq:model_general} as
\begin{subequations}
\begin{align}
H_0&=\sum_{i}\ef\,f^\dagger_if^{}_i-\sum_{\langle i,j\rangle}\left(\td\,c^\dagger_ic_j+\tf\,f^\dagger_if^{}_j+\hc\right)\nonumber\\
&-\sum_{\Nnn{i,j}}\left(\tdd\,c^\dagger_ic_j+\tff\,f^\dagger_if^{}_j+\hc\right)\\
&-\sum_{\Nnnn{i,j}}\left(\tddd\,c^\dagger_ic_j+\tfff\,f^\dagger_if^{}_j+\hc\right),\nonumber\\
H_{\rm hyb}&=\sum_{\alpha=x,y,z}\sum_{\langle i,j\rangle_\alpha}\!\!\left[\i V c_i^\dagger\sigma_\alpha f^{}_j\!+\!\i V f_i^\dagger\sigma_\alpha c_j\!+\!\hc\right],\\
H_{\rm int}&=\sum_iU\,f^\dagger_{i\uparrow}f^{}_{i\uparrow}\,f^\dagger_{i\downarrow}f^{}_{i\downarrow}\,,\label{eq:Hint}
\end{align}\label{eq:model}%
\end{subequations}
where $i,j\in\{1,\dots,N\}$ label the lattice sites. $\langle i,j\rangle$, $\Nnn{i,j}$, and $\Nnnn{i,j}$ denote pairs of nearest (\nn), next-to-nearest (\nnn), and next-to-next-to-nearest neighbors (\nnnn). The notation $\langle i,j\rangle_\alpha$ stands for a nearest-neighbor bond in the $\alpha$-direction and $\sigma_\alpha$ are the Pauli matrices in spin space. The annihilation (creation) operators $c_i$, $f_i$ ($c_i^\dagger$, $f_i^\dagger$) for the conduction $d$ electrons and $f$ electrons, respectively, are spinors $c_i=(c_{i\uparrow},c_{i\downarrow})^t$ and $f_i=(f_{i\uparrow},f_{i\downarrow})^t$. 
The particular form of the hybridization has been chosen in order to be odd under parity and to obey cubic as well as time-reversal symmetry.

We note that the model \eqref{eq:model} ignores the complicated multiplet structure of the $d$  and $f$ orbitals usually encountered in real Kondo insulators such as \smb. 
But importantly, the topological properties of cubic Kondo insulators do not depend on the particular shape of the orbitals or the precise form of the hopping and hybridization matrix elements. Instead, they follow directly from the points with band inversion, as we discuss in the course of this paper.
However, the multiplet structure may be important for identifying possible topological phases of specific materials as the orbital degeneracy can prevent the exchange of parity eigenvalues between valence and conduction bands at certain high-symmetry points (\hsp s). For example, the cubic symmetry enforces a twofold degeneracy of the relevant $\Gamma_8$ and $e_{\rm g}$ orbitals at the \ptt{\Gamma} and \ptt{R} points in \smb, such that the parity eigenvalue is always positive at those points.\cite{alexandrov_cubic_2013} 
In this paper, we do not consider such material-specific questions but instead we focus on the universal topological properties that are in principle possible in the presence of cubic symmetry.

The bandwidth of the $f$ electrons is much smaller than the bandwidth of the conduction electrons and we therefore assume that $|t_f|\ll |t_d|$ and similar relations hold for second- and third-neighbor hopping amplitudes. The hybridization is characterized by the parameter $V$ for which we typically use $|V|\lesssim |t_d|$. Throughout the whole paper, we choose $t_d$ to be the unit of energy, $\td=1$, and assume half filling.

\subsection{Noninteracting band structure}\label{sec:band structure}
We first comment on the noninteracting model. Thus, we assume $H_{\rm int}=0$ and analyze the Hamiltonian
\begin{align}
H_{\rm ni}&=H_0+H_{\rm hyb}\,.\label{eq:model_non-int}
\end{align}
For periodic boundary conditions, we can perform a Fourier transform which leads to
\begin{equation}
H_{\rm ni}=\sum_{\bk}\Psi^\dagger (\bk)\,h(\bk)\,\Psi(\bk)\,.
\end{equation}
Here, we defined the 4-spinor $\Psi(\bk)=\left(c_\uparrow,c_\downarrow,f_\uparrow,f_\downarrow\right)^t$ and the $4\times4$ Bloch matrix $h(\bk)$:\begin{equation}
\begin{split}
h(\bk)&=h_d(\bk)\,\tfrac{\E+\tau_z}{2}+h_f(\bk)\,\tfrac{\E-\tau_z}{2}+\Phi(\bk)\,\tau_x\\
&=\begin{pmatrix}h_d(\bk)&\Phi(\bk)\\\Phi(\bk)&h_f(\bk)\end{pmatrix}.
\label{eq:h}
\end{split}
\end{equation}
The Pauli matrices in orbital space are denoted by $\tau_i$ and $\E$ is the $2\times 2$ identity matrix. The dispersion of the $d$  and $f$ electrons is described by $h_d({\bs k})$ and $h_f({\bs k})$, respectively, and the hybridization by the matrix $\Phi({\bs k})$ as follows:
\begin{subequations}
\begin{align}
h_d(\bk)&=\left[-2\td\, c_1(\bk)-4\tdd\, c_2(\bk)-8\tddd\, c_3(\bk)\right]\E \,,\label{eq:hd}\\
h_f(\bk)&=\left[\ef-2\tf c_1(\bk)-4\tff c_2(\bk)-8\tfff c_3(\bk)\right]\E \,,\label{eq:hf}\\
\Phi(\bk)&=-2V\left(s_x\sigma_x+s_y\sigma_y+s_z\sigma_z\right) \,.\label{eq:Phi}
\end{align}\label{eq:h_d_f_Hyb}%
\end{subequations}%
Above, we used the definitions
\begin{subequations}
\begin{align}
c_1(\bk)&:=c_x+c_y+c_z\,,\\
c_2(\bk)&:=c_{xy}+c_{yz}+c_{zx}\,,\\
c_3(\bk)&:=c_xc_yc_z\,,
\end{align}%
\end{subequations}%
where $c_\alpha=\cos(k_\alpha)$, $c_{\alpha\beta}=c_\alpha c_\beta$, and $s_\alpha=\sin(k_\alpha)$ for $\alpha,\beta=x,y,z$. Note that the hybridization matrix \eqref{eq:Phi} is an odd function of $\bk$, $\Phi(\bk)=-\Phi(-\bk)$, but in order to preserve time-reversal symmetry, it also couples the physical spin of the electron. These properties are crucial for realizing a time-reversal-invariant topological Kondo insulator,\cite{dzero_topological_2010} as we will discuss in more details in the following.

Diagonalizing the Bloch matrix~\eqref{eq:h} yields the energy eigenvalues
\begin{equation}
E_\pm=\frac{E_d+E_f}{2}\pm\sqrt{\left(\frac{E_d-E_f}{2}\right)^2+E_{\rm hyb}^2}\,,
\label{eq:Epm}
\end{equation}
where we suppressed the $\bk$-label in the interest of better readability. Each band is twofold degenerate because of the combination of time-reversal and inversion symmetry. $E_d$, $E_f$, and $E_{\rm hyb}$ are the eigenvalues of $h_d(\bk)$, $h_f(\bk)$, and $\Phi({\bk})$, respectively, and are given by
\begin{subequations}
\begin{align}
E_d(\bk)&=-2\td c_1(\bk)-4\tdd c_2(\bk)-8\tddd c_3(\bk)\,,\\
E_f(\bk)&=\ef-2\tf c_1(\bk)-4\tff c_2(\bk)-8\tfff c_3(\bk)\,, \\
E_{\rm hyb}(\bk)&=-2V\sqrt{s_x^2+s_y^2+s_z^2}\,.
\end{align}
\end{subequations}
For future use, we also define the weight of the $d$ orbitals for a state vector $u$ as
\begin{align}
w(u):=u^\dagger\,\frac{\E+\tau_z}{2}\,u\,.
\end{align}
We say that $u$ has $d$ character ($f$ character) if $w(u)=1$ [$w(u)=0$]. We will also use the shorthand notation
\begin{equation}
w_a(\bk):=w[u_a(\bk)]\,,
\label{eq:wa}
\end{equation}
where $u_a(\bk)$ is the state of band $a$ at momentum $\bk$.

\begin{figure}
\centering
\figlabel{\includegraphics[width=.4\textwidth]{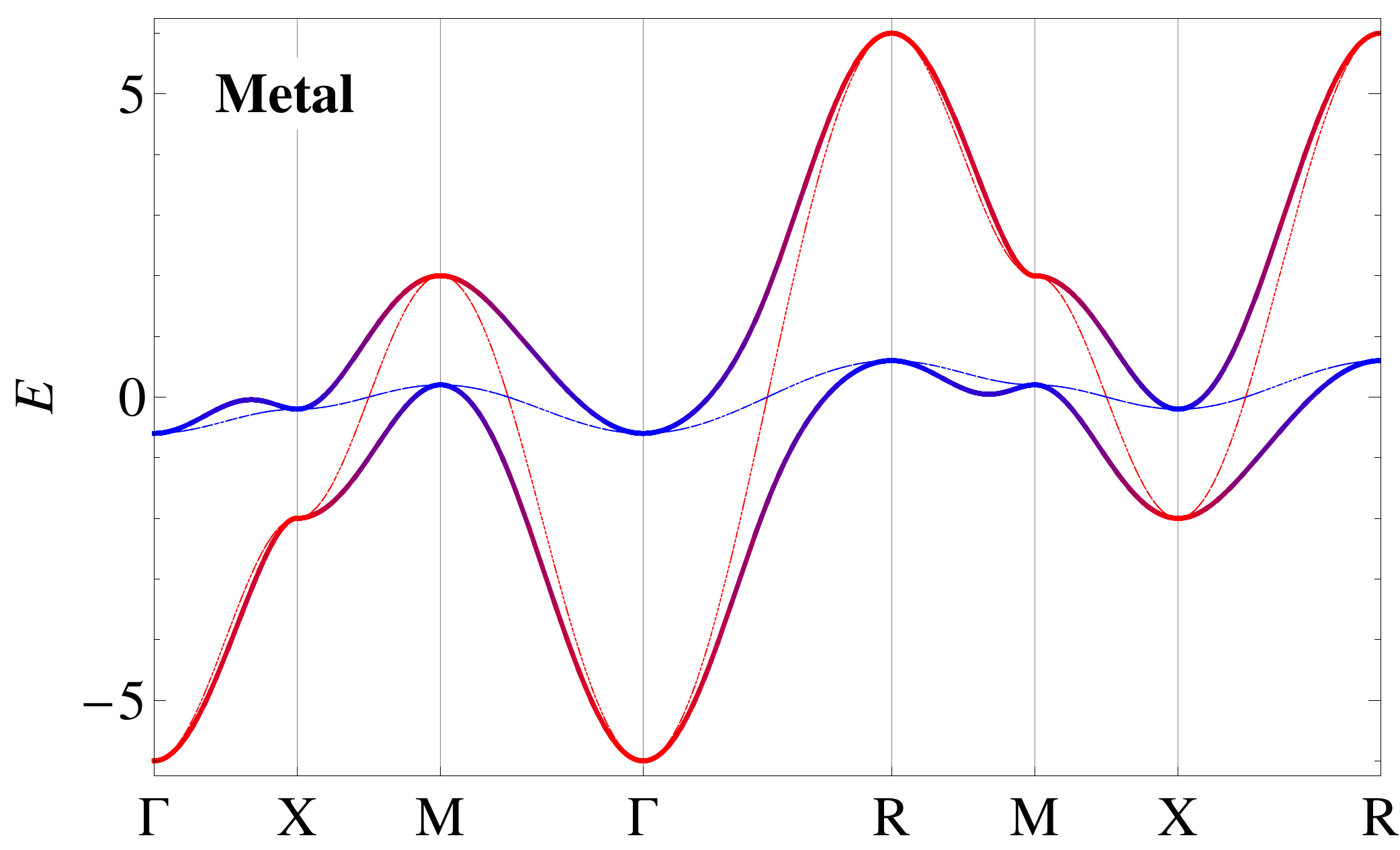}}{a}{0,0}\\[2mm]
\figlabel{\includegraphics[width=.4\textwidth]{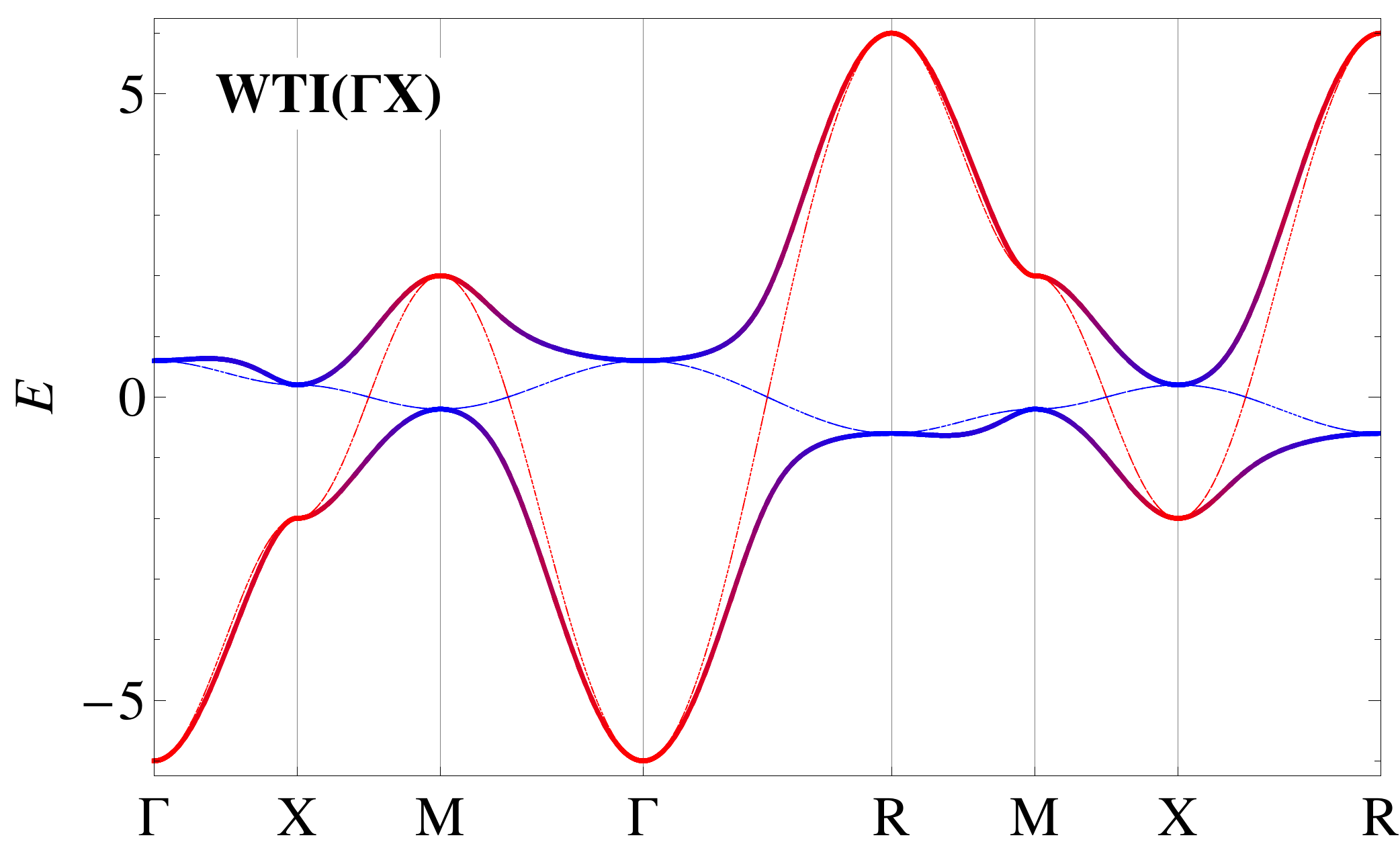}}{b}{0,0}
\caption{(Color online) Energy spectrum of the model~\eqref{eq:model_non-int} for $\td=1$, $|\tf|=0.1$, $\ttt=\tttt=0$, $V=0.5$, and $\ef=0$. The difference between the metallic (a) and insulating phase (b) is the sign of $\tf$: It is positive for the metallic and negative for the insulating phase. The shown insulating phase is \wti{\Gamma X} (see \tab{tab:invariants}).
The color shows $w(\bk)$; red (light gray) and blue (dark gray) denote $d$  and $f$ character, respectively. The thin dashed lines show the bare energies of $d$  and $f$ bands for the same parameters but with vanishing hybridization, $V=0$. The relatively large value for $V$ is chosen for a better visibility of the hybridization gap.}
\label{fig:energy_phases1}
\end{figure}

Figure~\ref{fig:energy_phases1} illustrates two exemplary band structures for which the narrow $f$ band lies within the conduction band. For these examples, we included only \nn hopping and in both cases the nonzero hybridization opens a direct gap. However, only if the sign of $t_f$ is opposite to the sign of $t_d$, also an indirect gap opens at half filling (and a weak topological insulator is found for the parameters of \fig{fig:energy_phases1}). Instead, if the signs are the same, the bands overlap and a metallic phase results. If additional further-neighbor hoppings are considered, insulating phases are possible if the ratios between first, second and third-neighbor hoppings for $f$ electrons are similar to the corresponding ratios for $d$ electrons, $\tdd/\td\approx\tff/\tf$ and $\tddd/\td\approx\tfff/\tf$. We will therefore assume $\tff=\tf (\tdd/\td)$ and $\tfff=\tf (\tddd/\td)$ in the following. 

\begin{figure}
\centering
\figlabel{\includegraphics[width=.4\textwidth]{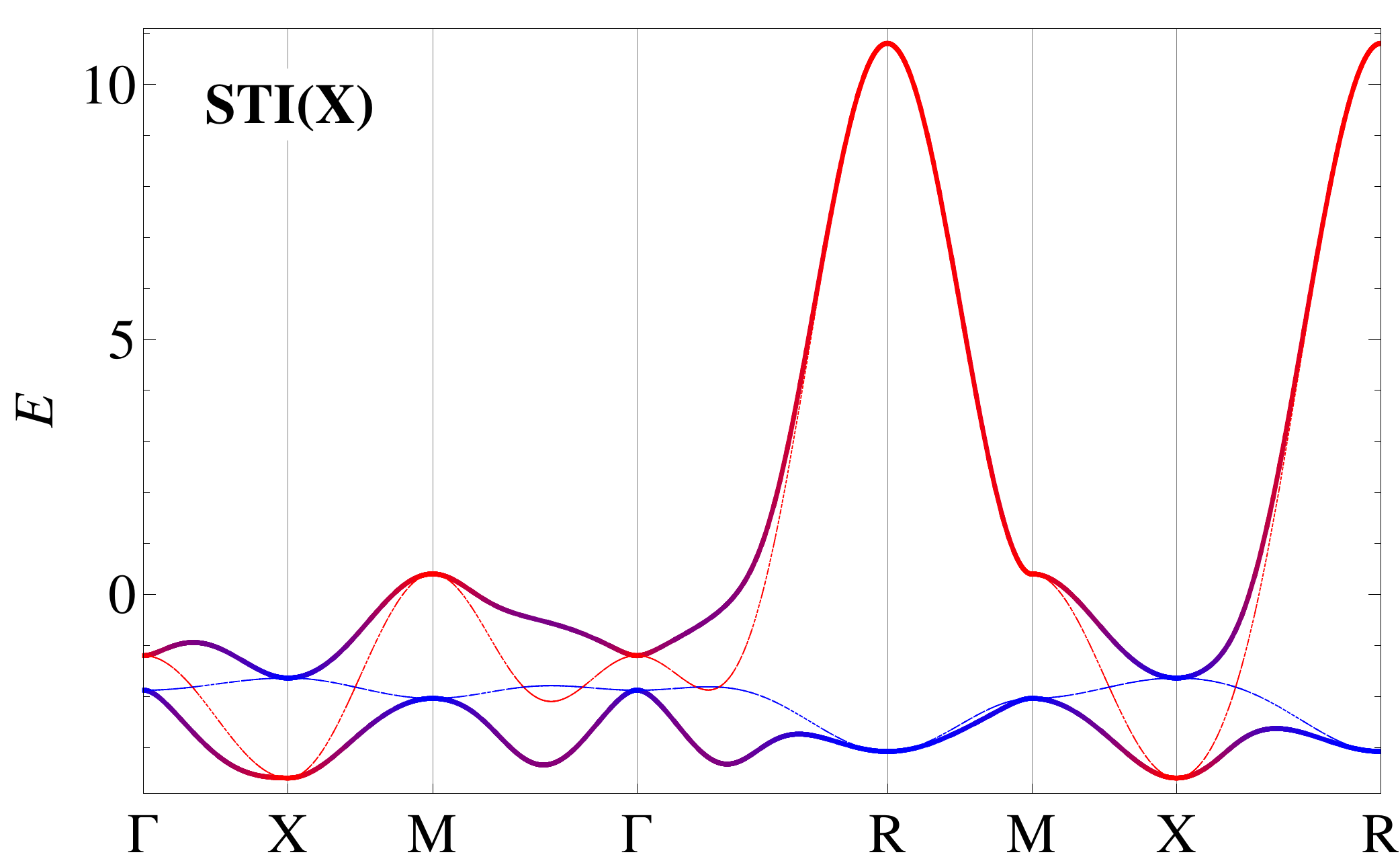}}{a}{0,1mm}\\[2mm]
\figlabel{\includegraphics[width=.4\textwidth]{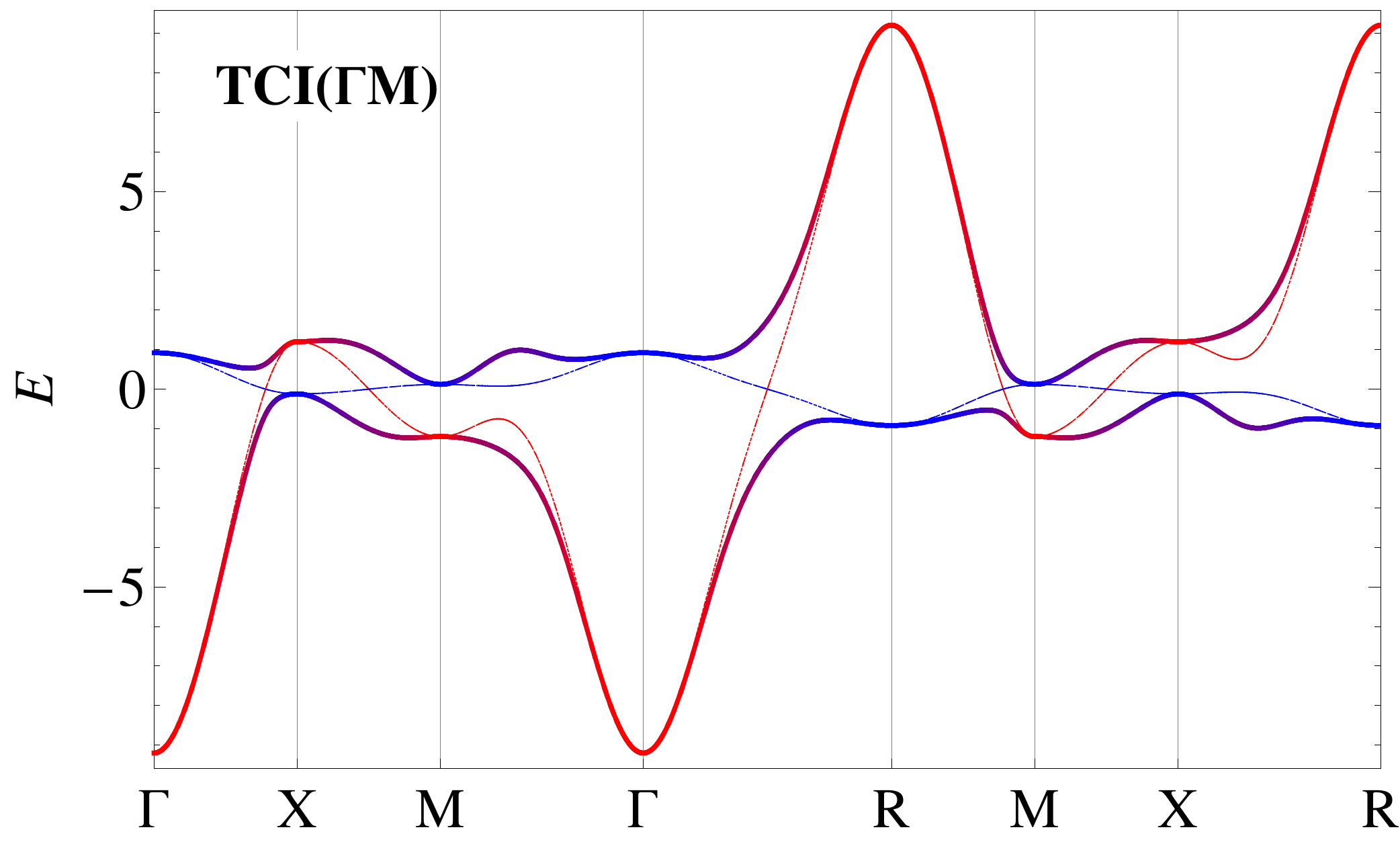}}{b}{0,1mm}
\caption{(Color online) Energy spectrum of the model~\eqref{eq:model_non-int} for different topological phases. The phase \sti{X} (a), which is relevant for \smb, requires \nnn\ hoppings, and the phase \tci{\Gamma M} (b) can only be realized when $\tttt>0.25\tt$.
For both plots, we chose the parameters $\td=1$, $\tf=-0.1$, and $V=0.5$. The remaining parameters are $\ttt=-0.4\tt$, $\tttt=0$, and $\ef=-2$ for \sti{X}; and $\ttt=0$, $\tttt=0.3\tt$, and $\ef=0$ for \tci{\Gamma M}.
The color coding and the different lines follow the same convention as in \fig{fig:energy_phases1}.}
\label{fig:energy_phases2}
\end{figure}

Figure~\ref{fig:energy_phases2} shows two more examples of band structures of gapped bulk phases realized in our model, which we will discuss more closely in the course of this paper. The respective phases are labeled by high-symmetry points with band inversions, according to a convention discussed in \Sec{sec:band-inversions} and \tab{tab:invariants}. In Fig.~\ref{fig:energy_phases2}(a), we show the band structure of the phase \sti{X}, which has a band inversion at the three \ptt{X} points and realizes a strong topological insulator. From an experimental point of view this phase is especially interesting, as it is expected that \smb\ also has band inversions at the \ptt{X} points.\cite{takimoto_smb6_2011,lu_correlated_2013,alexandrov_cubic_2013,dzero_new_2013} In \fig{fig:energy_phases2}(b), we show the band structure of the phase \tci{\Gamma M}, which has band inversions at the $\Gamma$ and the three M points. All the $\zz$ invariants are trivial in this phase but it has nontrivial mirror Chern numbers. It thus realizes a topological crystalline insulator.\cite{fu_topological_2011,hsieh_tci_2012,tanaka_experimental_2012,dziawa_topological_2012,xu_observation_2012}

Note that band energies, such as shown in Figs.~\ref{fig:energy_phases1} and \ref{fig:energy_phases2}, are given in accordance with Eq.~\eqref{eq:model_non-int}. Thus, the Fermi level at half filling is located at different energies for different choices of band parameters.

\subsection{Renormalized band structure from interactions}
\label{sec:ren}
Interactions among the $f$ electrons can have various effects on the system, stabilizing, e.g., magnetically ordered states, heavy or non-Fermi liquids, as well as unconventional superconductors. Here, we restrict our analysis to the situation typically found in Kondo insulators: Interactions strongly renormalize the band parameters but the low-energy excitations are still described by well-defined Fermi-liquid quasiparticles. Thus, we treat the interactions in the quasiparticle approximation to the periodic Anderson model,\cite{Read:1983b,Coleman:1984,Rice:1985b,Yamada:1986,kotliar_new_1986} assuming a ${\bs k}$-independent self-energy for the $f$ electrons of the Fermi-liquid type $\Sigma_f(\omega)=a+b\omega+\mathcal{O}(\omega^2)$. The Fermi-liquid quasiparticles in such a state are then accurately described by a noninteracting Hamiltonian with renormalized parameters, which depend on the interaction $U$ and the noninteracting band parameters. Specifically, all $f$-electron hopping amplitudes are renormalized by $\tf \to z^2\tf$, the hybridization by $V\to zV$, and the onsite-energy of $f$ electrons by $\ef \to \ef + \lambda$, where the parameters $z$ and $\lambda$ are related to the coefficients in the expansion of the self-energy by $z=(1-b)^{-1/2}$ and $\lambda=a/(1-b)$. The dependence of $z$ and $\lambda$ on the band parameters and $U$ can be studied nonperturbatively by various methods, including dynamical mean-field theory, Gutzwiller projected variational wave-functions, or slave-particle representations. 

In this work, we use the Kotliar-Ruckenstein slave-boson scheme in the mean-field approximation\cite{kotliar_new_1986} to obtain the quasiparticle Hamiltonian. The topological properties can then be analyzed in the same way as those of the noninteracting Hamiltonian. An interesting question, which we will address in \Sec{sec:interactions}, is, whether interactions are capable of driving topological phase transitions. We will demonstrate that this is indeed the case: Depending on the noninteracting band parameters, increasing $U$ can drive a transition from a topological to a trivial  or from a trivial to a topological phase.

\section{Topological classification}\label{sec:top}
\subsection{Summary}\label{sec:top_sum}
In the following sections, we provide a complete topological classification of the time-reversal-invariant gapped phases obtained in the quasiparticle approximation to the interacting model~\eqref{eq:model_general} (see \Sec{sec:ren}). For notational simplicity, the band parameters are denoted with their bare values, but they can equally well be understood as the renormalized values in the interacting model. Depending on the relative magnitude of \nn, \nnn, and \nnnn\ hopping for $d$  and $f$ electrons, as well as the onsite potential and the interaction between $f$ electrons, a metallic, trivial insulating or one of several different topological phases is realized. An overview of the different insulating phases is given in \tab{tab:invariants}.

In \Sec{sec:top_inv}, we characterize these different phases using different symmetry-protected topological invariants and in \Sec{sec:surface}, we illustrate how the surface state properties can be understood from knowledge of these invariants.

\subsubsection{Topological phases from band inversions}\label{sec:band-inversions}
Because our model respects the inversion symmetry of the cubic lattice, we first discuss how the different phases are distinct by the inversion eigenvalues at \hsp s in the Brillouin zone. The odd-parity property of the hybridization function~\eqref{eq:Phi} implies that the two orbitals ($f$  and $d$ electrons) have opposite parity. Hence, the inversion operator is represented as $I=\tau_z$ and the Bloch Hamiltonian~\eqref{eq:h} satisfies
\begin{equation}
I^{-1}h(-\bk)I=h(\bk)\,.
\end{equation}
Because of cubic symmetry, only four out of eight inversion symmetric momenta $\Gamma_i\in\{0,\pi\}^3$ (which coincide with the \trim) are independent, and, following standard notation, we denote them as $\pt{\Gamma}=(0,0,0)$, $\pt{X}\in\{(\pi,0,0),\,(0,\pi,0),\,(0,0,\pi)\}$, $\pt{M} \in\{(\pi,\pi,0),\,(0,\pi,\pi),\,(\pi,0,\pi)\}$, and $\pt{R}=(\pi,\pi,\pi)$. The three \ptt{X} points are equivalent for symmetry reasons, as are the three \ptt{M} points. At these \hsp s, we have a vanishing hybridization, $\Phi(\Gamma_i)=0$, and therefore each Kramers pair has pure $d$  or $f$ character. 

We define the points with band inversion as those \hsp s where the occupied states have $d$ character (instead of $f$ character) and label the corresponding phase with the \hsp (s), at which the band inversion occurs (see \tab{tab:invariants}). If there are more than two of those points, we instead list the points where the occupied states have $f$ character and denote the respective phase with a bar. With this convention, each phase is labeled with at most two \hsp s with a band inversion. In total, there are 16 different phases with different occupations at the \hsp s, but always two phases are related to each other by inverting the occupations at all the \hsp s, which can be achieved by flipping the sign of all the hopping amplitudes. In \tab{tab:invariants}, we list the remaining eight independent phases.
\begin{table}
\caption{Topological invariants and mirror Chern numbers for different insulating phases grouped into band insulator (\bi), topological crystalline insulator (\ttci), weak topological insulator (\wwti), and strong topological insulator (\ssti). The different phases are labeled by the \hsp s with an occupied $d$ band. 
For an inversion of $f$ electrons at the respective points, the mirror Chern numbers must be multiplied by $-1$, and we will denote the respective phase by a bar, e.g., \stii{M}.
The second column shows, which hopping amplitudes of our model are required to be nonzero in order to create the respective phases. For example, the phase \tci{\Gamma M} only occurs when including \nnnn\ hopping.}
\label{tab:invariants}
\begin{ruledtabular}
\begin{tabular}{ccccccc}
phase&\begin{tabular}{@{}c@{}}required\\[-1.8mm]hopping\end{tabular}&$(\nu_0;\nu_1,\nu_2,\nu_3)$&$C^+_{k_z=0}$&$C^+_{k_z=\pi}$&$C^+_{k_x=k_y}$\\ \hline
\bi&none&(0;0,0,0)&0&0&0\\\hline
\tci{\Gamma M}&$\tttt$&(0;0,0,0)&2&-2&0\\ \hline
\wti{\Gamma X}&$\tt$&(0;1,1,1)&-1&1&0\\ 
\wti{\Gamma R}&$\ttt$&(0;1,1,1)&1&1&2\\ \hline
\sti{\Gamma}&$\tt$&(1;0,0,0)&1&0&1\\
\sti{X}&$\tt,\ttt$&(1;1,1,1)&-2&1&-1\\
\sti{M}&$\tt,\ttt$&(1;0,0,0)&1&-2&-1\\
\sti{R}&$\tt$&(1;1,1,1)&0&1&1
\end{tabular}
\end{ruledtabular}
\end{table}
\subsubsection{Phase diagrams}\label{sec:phase-diagrams}
\begin{figure}
\centering
\includegraphics[width=.4\textwidth]{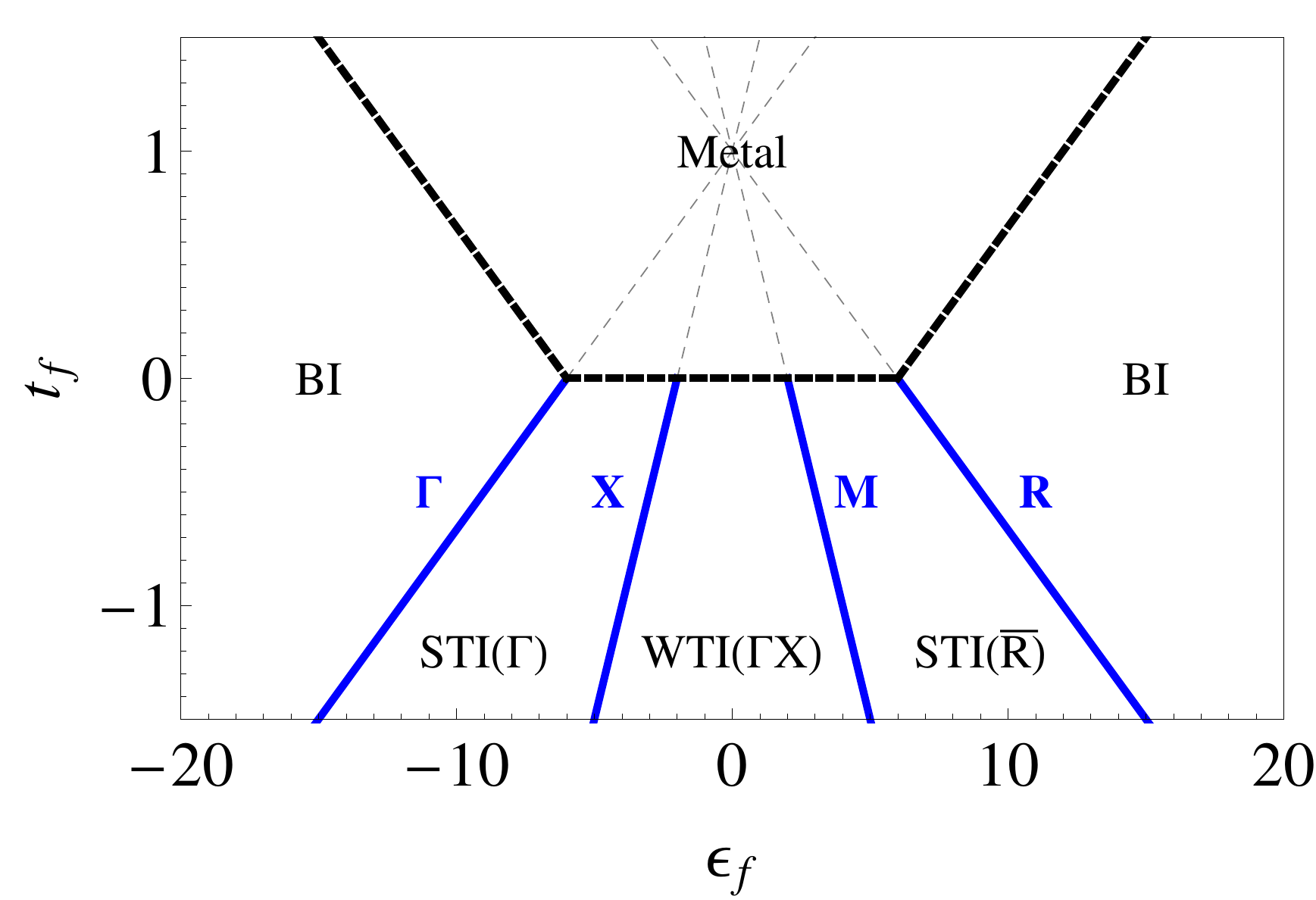}
\caption{(Color online) Phase diagram of the noninteracting model~\eqref{eq:model_non-int} for $\td=1$ and $\ttt=\tttt=0$. The value of the hybridization (as long as $V\neq0$) does not influence the phase diagram.
The labels for the different phases correspond to the convention of \tab{tab:invariants}. In addition, there is a metallic phase where $d$  and $f$ bands overlap for $\tf>0$. 
The thick blue lines show phase transitions and are labeled according to the \hsp\ at which the energy gap closes. These are continued inside the metallic phase as thin, dashed gray lines, where the direct bandgap closes at the respective points. The thick, dashed, black lines show where the bands start to overlap.}
\label{fig:phases_tf_ef}
\end{figure}

We now discuss how the different trivial and topological phases depend on the tight-binding parameters. First, we consider the case of vanishing \nnn\ and \nnnn\ hopping, $\tdd=\tff=\tddd=\tfff=0$. Then, the two remaining parameters are $\tf$ and $\ef$. The phase diagram for this case is shown in \fig{fig:phases_tf_ef}. 
We observe that with only \nn\ hoppings, a single band inversion is possible at the \ptt{\Gamma} or the \ptt{R} point but not at the \ptt{X} or \ptt{M} points. At the phase transitions, the energy gap closes. This must happen at a \trim, as these are the only points where there is no hybridization gap. From the general expression of the band energies~\eqref{eq:Epm}, the condition for a gap closing at $\Gamma_i$ is obtained as
\begin{equation}
{\tf}(\Gamma_i)=\td+\frac{\ef}{2c_1(\Gamma_i)}\,.
\label{eq:tfGammai}
\end{equation}
For the four different \hsp s, \eq{eq:tfGammai} reads as
\begin{subequations}
\begin{align}
\tf(\pt{\Gamma})&=\td+\tfrac16 \ef\,,\\
\tf(\pt{X})&=\td+\tfrac12 \ef\,,\\
\tf(\pt{M})&=\td-\tfrac12 \ef\,,\\
\tf(\pt{R})&=\td-\tfrac16 \ef\,.
\end{align}%
\label{eq:tf}%
\end{subequations}%
In \fig{fig:phases_tf_ef}, these are the lines between the different topological regions for $\tf<0$. The transition from insulating to metallic behavior at $\tf=0$ is not associated with a closing of the direct band gap, but by the closing of the indirect gap. The lines given in \eq{eq:tf} therefore extend also into the metallic region at $t_f>0$.

\begin{figure}
\centering
\figlabel{\includegraphics[width=.4\textwidth]{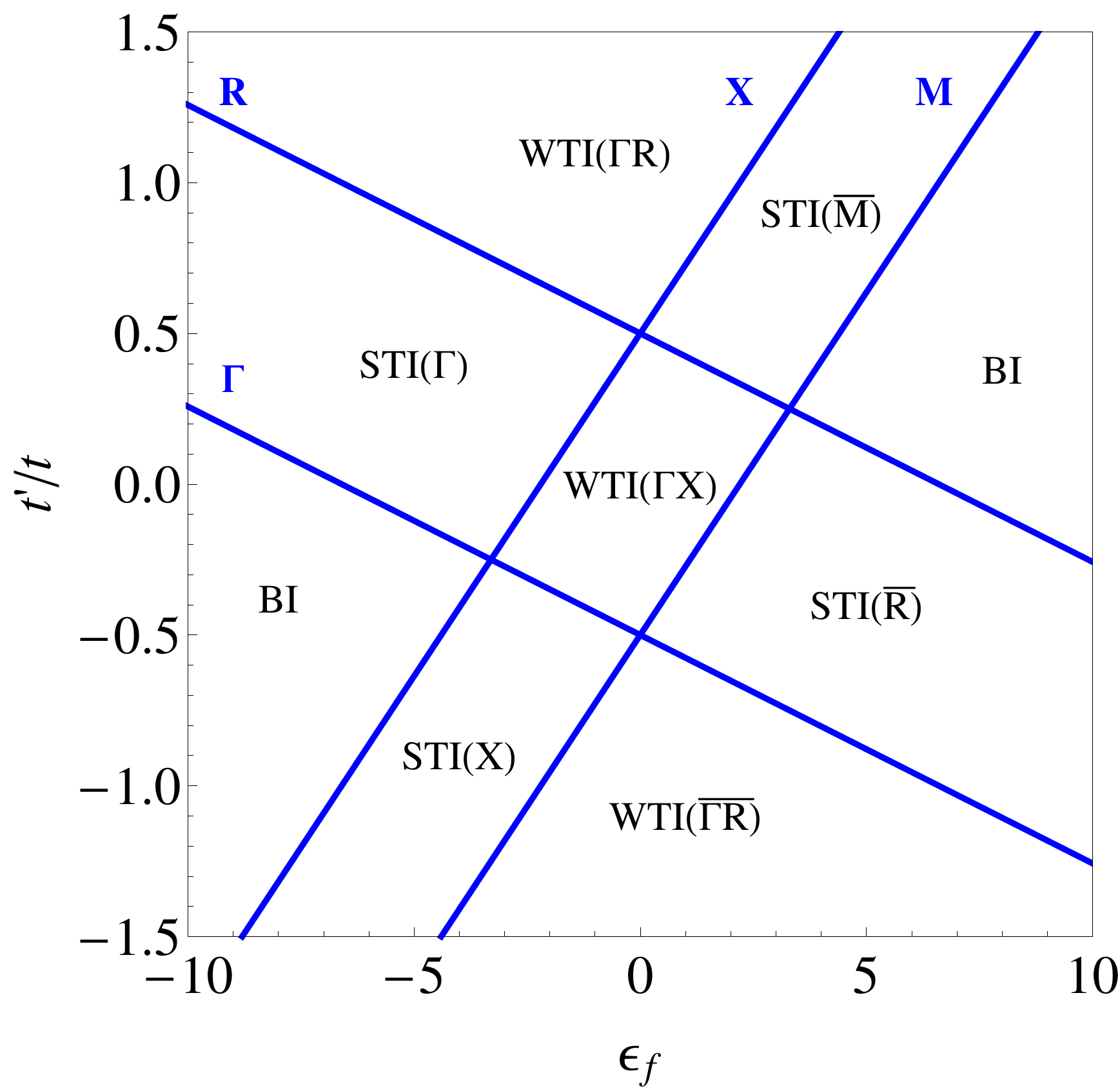}}{a}{0,0}
\\[5mm]
\figlabel{\includegraphics[width=.4\textwidth]{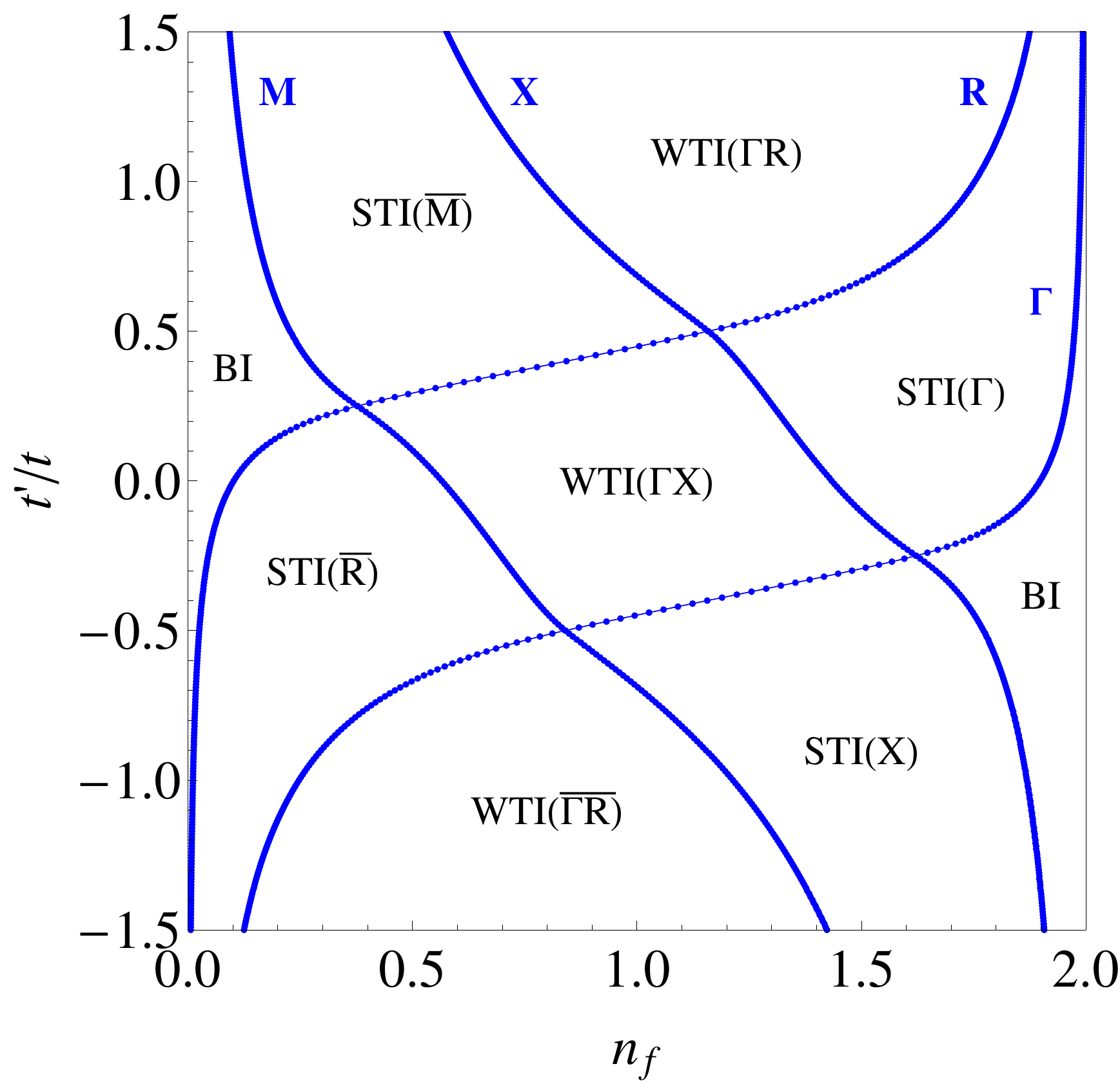}}{b}{0,0}
\caption{(Color online) Phase diagram of the noninteracting model~\eqref{eq:model_non-int} for $\td=1$ and $\tf=-0.1$, as a function of the \nnn\ hopping $\tdd$ and the chemical potential $\ef$ (a), and of the filling fraction $\nf$ of $f$ orbitals (b). The ratios of \nn\ and \nnn\ hopping are equal for $d$  and $f$ electrons, $\tdd/\td=\tff/\tf$ and we assume vanishing \nnnn\ hopping, $\tttt=0$. Although the value of the hybridization (as long as $V\neq0$) does not influence the $\ttt$-$\ef$ phase diagram, it influences the relation between $\ef$ and $\nf$; the $(\ttt/\tt)$-$\nf$ diagram was created using $V=0.5$. 
The labeling conventions are the same as in \fig{fig:phases_tf_ef}.}
\label{fig:phases_tff_ef}
\end{figure}

\medskip

Now we want to analyze how the situation changes when we allow for nonzero \nnn\ hopping amplitudes. 
As discussed in \Sec{sec:band structure}, we consider arbitrary \nnn\ hopping which obeys the condition $\tdd/\td=\tff/\tf$. We are interested in the insulating region, so we fix $\tf<0$ and $V\neq0$. The resulting phase diagram for this choice is shown in \fig{fig:phases_tff_ef}(a) in the $(\ttt/\tt)$-$\ef$ plane and in (b) in the $(\ttt/\tt)$-$\nf$ plane, where $\nf$ denotes the occupation of the $f$ orbitals:
\begin{equation}
\nf=\sum_a\frac{1}{(2\pi)^3}\int_{\text \bz }\d^3 k\,[1-w_a(\bk)]\,.
\end{equation}
Here, the sum is taken over all occupied bands and we integrate the weight function~\eqref{eq:wa} over the Brillouin zone (\bz). Because $0\leq w\leq 1$ and because we always consider two occupied bands, the $f$-orbital filling must satisfy $0\leq \nf\leq 2$.

In the $(\ttt/\tt$)-$\ef$ diagram, we can again analytically obtain the phase transition lines by considering the energy~\eqref{eq:Epm} at the \hsp s. The general condition for the gap closing is
\begin{equation}
(\ttt/\tt)({\Gamma_i})=-\frac{c_1(\Gamma_i)}{2c_2(\Gamma_i)}-\frac{\ef}{4(\td-\tf)c_2(\Gamma_i)}\,,
\end{equation}
which leads to the following four different lines shown in \fig{fig:phases_tff_ef}:
\begin{subequations}
\begin{align}
(\ttt/\tt)(\pt{\Gamma})&=-\tfrac12-\frac1{12(\td-\tf)}\,\ef\,,\\
(\ttt/\tt)(\pt{X})&=+\tfrac12+\frac1{4(\td-\tf)}\,\ef\,,\\
(\ttt/\tt)(\pt{M})&=-\tfrac12+\frac1{4(\td-\tf)}\,\ef\,,\\
(\ttt/\tt)(\pt{R})&=+\tfrac12-\frac1{12(\td-\tf)}\,\ef\,.
\end{align}
\end{subequations}
Because the relation between $\ef$ and $\nf$ is nonlinear, these lines map onto complicated curves in the $(\ttt/\tt)$-$\nf$ diagram. Note, however, that the topology of the phase diagram is the same in both cases.

If the $f$ orbitals are half filled, $\nf=1$, the narrow $f$ band necessarily lies in the middle of the $d$ band, which for almost all choices of parameters leads to two \hsp s with band inversion. This explains why a weak topological insulator phase is favored in this regime, which is consistent with the finding in \Ref{dzero_topological_2010}.

\medskip

\tab{tab:invariants} also lists the hopping amplitudes which are required to be nonzero for the eight different phases. All phases can be realized with only \nn\ and \nnn\ hopping, except for the \tci{\Gamma M} phase. This phase, which is characterized by vanishing $\zz$ invariants but has nonzero mirror Chern numbers, can only be realized when $\tttt>0.25\tt$. The bulk and surface band structure are shown in Figs.~\ref{fig:energy_phases2}(b) and \ref{fig:surface3}, respectively, but the phase is not present in any of the phase diagrams where we always assumed $\tttt=0$.

\subsection{Topological invariants}
\label{sec:top_inv}
We demonstrate that the band inversions uniquely define a set of topological invariants. These are the four $\zz$ invariants $(\nu_0;\nu_1,\nu_2,\nu_3)$ as well as three mirror Chern numbers $C_{k_z=0}^+$, $C^+_{k_z=\pi}$ and $C_{k_x=k_y}^+$ associated with three independent mirror planes.

\subsubsection{$\zz$ invariants}\label{sec:time-reversal}
Our model~\eqref{eq:model_non-int} belongs to the class AII in the notation of Altland and Zirnbauer and is therefore characterized by a $\zz$ topological invariant,\cite{altland_nonstandard_1997,schnyder_classification_2008} which is protected by time-reversal symmetry and particle conservation. This is the strong topological index $\nu_0$. In addition, because of (discrete) translation symmetry, we can also define three weak topological $\zz$ indices ($\nu_1$, $\nu_2$, $\nu_3$).\cite{fu_topological3D_2007} These are the $\zz$-invariants of the associated two-dimensional systems (in the same symmetry class) obtained by fixing one of the momentum components at a time-reversal-invariant value $k_i=\pi$.

In the presence of inversion symmetry, these $\zz$ invariants are directly related to the inversion eigenvalues of occupied states at the \trim.\cite{fu_topological_2007} For example, the strong topological index is given by
\begin{subequations}
\begin{equation}
(-1)^{\nu_0}=\prod_{j=1}^8 \prod_{a} \xi\left[u_a(\Gamma_j)\right]\,,
\end{equation}
where the second product runs over all occupied Kramers pairs and $\xi(u)=\pm1$ is the inversion eigenvalue of the state vector $u$, $I|u_a(\Gamma_i)\rangle=\xi(u_a(\Gamma_i))|u_a(\Gamma_i)\rangle$. Similarly, the weak indices $\nu_i$ ($i=1,2,3$) are defined by a product of parities on the planes $k_i=\pi$:
\begin{equation}
(-1)^{\nu_i}=\prod_{j;k_i=\pi} \prod_{a} \xi\left[u_a(\Gamma_j)\right]\,.
\end{equation}
\end{subequations}
Using cubic symmetry, these expressions simplify to
\begin{subequations}
\begin{align}
\allowdisplaybreaks[2]
\begin{split}
(-1)^{\nu_0}&=\prod_a \xi\left[u_a(\pt{\Gamma})\right]\,\xi^3\left[u_a(\pt{X})\right]\,\xi^3\left[u_a(\pt{M})\right]\,\xi\left[u_a(\pt{R})\right]\\
&=\prod_a \xi\left[u_a(\pt{\Gamma})\right]\,\xi\left[u_a(\pt{X})\right]\,\xi\left[u_a(\pt{M})\right]\,\xi\left[u_a(\pt{R})\right]\,,\label{eq:strong}
\end{split}\\
\begin{split}
(-1)^{\nu_i}&=\prod_a \xi\left[u_a(\pt{X})\right]\,\xi^2\left[u_a(\pt{M})\right]\,\xi\left[u_a(\pt{R})\right]\\
&=\prod_a \xi\left[u_a(\pt{X})\right]\,\xi\left[u_a(\pt{R})\right]\,.\label{eq:weak}
\end{split}
\end{align}
\label{eq:Z2}
\end{subequations}
Note that because of cubic symmetry, the weak indices are all equal, $\nu_1=\nu_2=\nu_3$. As we always calculate products over an even number of values $\pm 1$ for the topological invariants, the result is not changed by switching the parity of $d$ and $f$ orbitals ($I\rightarrow -I$). Moreover, the strong index only depends on the parity of the number of band inversions (it is nonzero for an odd number of band inversions and vanishes for an even number) and according to \eq{eq:weak}, the weak indices depend only on the inversion eigenvalues at \ptt{X} and \ptt{R}. It is thus apparent that knowledge of the $\zz$ invariants does not uniquely determine the band inversions (see \tab{tab:invariants}). Nevertheless, in the presence of cubic symmetry, this finer classification can be obtained from a topological invariant called the mirror Chern number,\cite{teo_surface_2008} which will be discussed next.

\subsubsection{Mirror Chern numbers}\label{sec:mirror}
\newsavebox{\figc}
\savebox{\figc}{%
\begin{tikzpicture}[scale=.6]
\node (a) at (-8,0){}; \node (b) at (-5.72,0){}; \node (c) at (-1.32,0){}; \node (e) at (3.96,0){}; \node (d) at (5,0){};
\draw[->] (a) -- (d) node[right]{$\ef$};
\draw[very thick,blue] (b)++(0,-.1) -- ++(0,.2) node [above] {\ptt{X}};
\draw[very thick,blue] (c)++(0,-.1) -- ++(0,.2) node [above] {\ptt{M}};
\draw[very thick,blue] (e)++(0,-.1) -- ++(0,.2) node [above] {\ptt{\Gamma}};
\node at ($(a)!.5!(b)+(0,.4)$) {\bi};
\node at ($(b)!.5!(c)+(0,.4)$) {\sti{X}};
\node at ($(c)!.5!(e)+(0,.4)$) {\wtii{\Gamma R}};
\draw[thick,red,->] (-5,0) ++(0,-.2) node[below]{A} -- ++(0,.2);
\draw[thick,red,->] (-3,0) ++(0,-.2) node[below]{B} -- ++(0,.2);
\draw[thick,red,->] (-.5,0) ++(0,-.2) node[below]{C} -- ++(0,.2);
\end{tikzpicture}}

\begin{figure}[t]
\centering
\figlabel{\usebox{\figc}}{a}{0,0}\\[3mm]
\figlabel{\includegraphics[width=.48\textwidth]{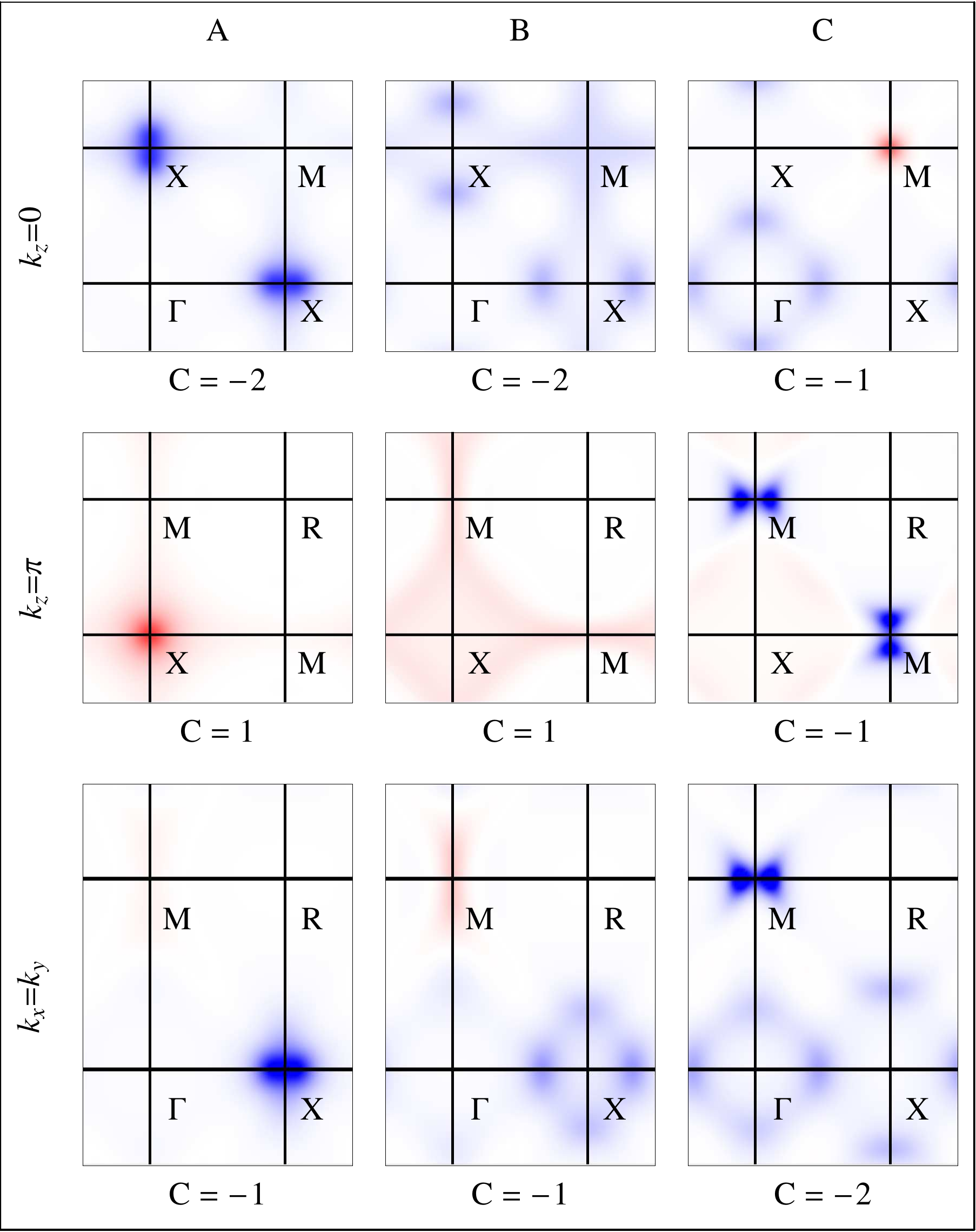}}{b}{1mm,-1mm}\\[2mm]
\includegraphics[width=.4\textwidth]{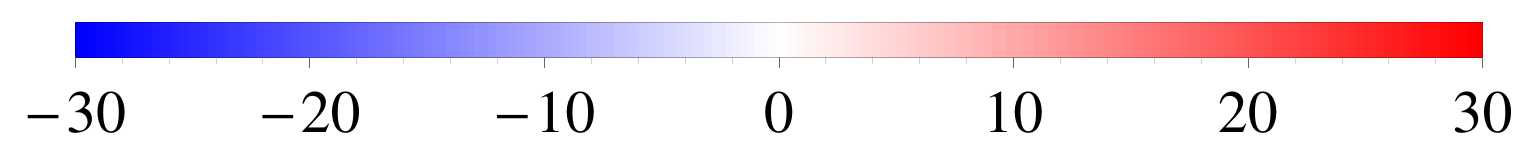}
\caption{(Color online) The drawing (a) shows the different phases and phase transitions for the chosen parameters and variable $\ef$, as well as the choices for $\ef$ in the different columns (A, B, C).\\
The plots in (b) show the Berry curvature~\eqref{eq:berry} for the three mirror-invariant planes $k_z=0$ (top), $k_z=\pi$ (middle), and $k_x=k_y$ (bottom). The parameters are $\td=1$, $\tf=-0.1$, $\ttt/\tt=-0.8$, $\tttt=0$, and $V=0.5$; the chemical potential $\ef$ varies from $\ef=-5$ (A) to $\ef=-3$ (B) and $\ef=-0.5$ (C). Close to the phase transitions, the Berry curvature is strongly peaked at the \hsp\ where the gap closes (columns A and C). At the phase transition it has a jump of $\pm1$ at the respective \hsp. Away from the phase transitions, the Berry curvature is delocalized (column B). Note that the Berry curvature only has twofold rotational symmetry around the \ptt{X} points (\ptt{M} points) on the $k_z=0$ and $k_x=k_y$ ($k_z=\pi$ and $k_x=k_y$) planes.
}
\label{fig:berry}
\end{figure}

\newsavebox{\figd}
\newcommand*{\dotrad}{0.15}
\tikzset{
  c/.style={every coordinate/.try}
}
\savebox{\figd}{%
\begin{tikzpicture}[scale=.5]

\coordinate (g) at (0,0); \coordinate (x1) at (5,0); \coordinate (x2) at (0,5); \coordinate (m3) at (5,5);
\begin{scope}[shift={($0.35*(-5,-5)$)}]
\coordinate (x3) at (0,0); \coordinate (m1) at (5,0); \coordinate (m2) at (0,5); \coordinate (r) at (5,5);
\end{scope}
\draw (g) node[above right]{\ptt{\Gamma}} -- (x1) node[below=1mm]{\ptt{X}} -- (m3) node[above left,name=b]{\ptt{M}} -- (x2) node[left=1mm]{\ptt{X}} -- cycle;
\draw (x3) node[below right,name=a]{\ptt{X}} -- (m1) node[right=1mm]{\ptt{M}} -- (r) node[below left]{\ptt{R}} -- (m2) node[above=1mm]{\ptt{M}} -- cycle;
\draw (g) -- (x3) (x1) -- (m1) (m3) -- (r) (x2) -- (m2);
\draw[dashed] (g) -- (m1) (x2) -- (r);
\filldraw[red] (g) circle (\dotrad) (x1) circle (\dotrad) (x2) circle (\dotrad) (x3) circle (\dotrad) (m1) circle (\dotrad) (m2) circle (\dotrad) (m3) circle (\dotrad) (r) circle (\dotrad);
\begin{scope}[every coordinate/.style={shift={($2*\dotrad*(1,0)$)}}]
\filldraw[blue] ([c]x1) circle (\dotrad) ([c]m2) circle (\dotrad);
\end{scope}
\begin{scope}[every coordinate/.style={shift={($2*\dotrad*(-1,0)$)}}]
\filldraw[blue] ([c]x1) circle (\dotrad) ([c]m2) circle (\dotrad);
\end{scope}
\begin{scope}[every coordinate/.style={shift={($2*\dotrad*(0,1)$)}}]
\filldraw[blue] ([c]x2) circle (\dotrad) ([c]m1) circle (\dotrad);
\end{scope}
\begin{scope}[every coordinate/.style={shift={($2*\dotrad*(0,-1)$)}}]
\filldraw[blue] ([c]x2) circle (\dotrad) ([c]m1) circle (\dotrad);
\end{scope}
\begin{scope}[every coordinate/.style={shift={($1.414*\dotrad*(1,1)$)}}]
\filldraw[blue] ([c]x3) circle (\dotrad) ([c]m3) circle (\dotrad);
\end{scope}
\begin{scope}[every coordinate/.style={shift={($-1.414*\dotrad*(1,1)$)}}]
\filldraw[blue] ([c]x3) circle (\dotrad) ([c]m3) circle (\dotrad);
\end{scope}
\filldraw[red] ($(a)!.5!(b)+(5,.5)$) circle (\dotrad) node[right=2mm]{$+1$};
\filldraw[blue] ($(a)!.5!(b)+(5,-.5)$) circle (\dotrad) node[right=2mm]{$-1$};
\end{tikzpicture}}

\begin{figure}[t]
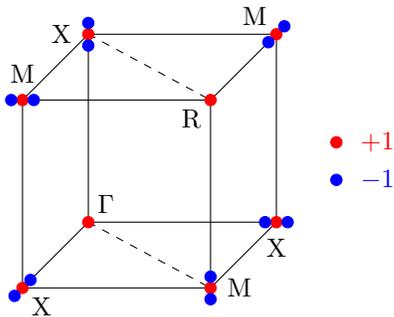

\centering
\usebox{\figd}
\caption{(Color online) Illustration of the relations \eqref{eq:mirror-rules} for the mirror Chern numbers: As the Berry flux at the \hsp s changes by $\pm 1$ when creating a band inversion at the respective point, any mirror Chern number can be obtained by summing up all the ``charges" $\pm 1$ lying in the considered mirror plane for those \hsp s with band inversion. The charges $\pm1$ are shown as red and blue dots. Note that for the \ptt{X} and \ptt{M} points, the charges are different for the different mirror planes, reflecting the smaller rotational symmetry group at those points. While there is a freedom to choose an overall sign for the different charges, their relative sign is fixed by the condition that the sum of all charges in any plane must vanish.}
\label{fig:charges}
\end{figure}

The definition of the mirror Chern number requires that the point-group symmetry contains mirror operations.\cite{teo_surface_2008} There is one invariant plane in momentum space for each mirror plane that is invariant under the mirror operation $M=IC_2$, where $I$ is the inversion and $C_2$ is a rotation by $\pi$ about an axis perpendicular to the mirror plane. Because $M^2=-1$ for spin-$\tfrac12$ particles, the state vectors in a mirror plane can be chosen to have eigenvalues $\pm\i$ under the mirror operation:
\begin{equation}
M\ket{u_a^{\pm}(\tilde \bk)}=\pm\i\ket{u_a^{\pm}(\tilde \bk)}\,,\label{eq:mirror_es}
\end{equation}
where the momentum $\tilde\bk$ lies in this mirror plane.
There, we can define the Berry connection of the band with mirror eigenvalue $\pm\i$,
\begin{equation}
\bm{\mathcal{A}}^\pm_a(\tilde\bk)=\bra{u_a^{\pm}(\tilde\bk)}\nabla_{\tilde\bk}\ket{u_a^{\pm}(\tilde\bk)}\,,\label{eq:berry_conn}
\end{equation}
and the corresponding Berry curvature
\begin{equation}
\mathcal{F}^\pm_{a}(\tilde\bk)=\frac{\partial\mathcal{A}^\pm_{a,2}(\tilde\bk)}{\partial\tilde k_1}-\frac{\partial\mathcal{A}^\pm_{a,1}(\tilde\bk)}{\partial\tilde k_2}\,\label{eq:berry_curv}
\end{equation}
where $\nabla_{\tilde\bk}$ denotes the gradient in the mirror plane and we follow the conventions of \Ref{suzuki_chern_2005}.

Then, we define the mirror Chern number associated with a particular mirror operation as the Chern number $C^+$ of the occupied states with eigenvalues $+\i$:
\begin{equation}
C^+=\frac{1}{2\pi\i}\int_{\overline{\rm \bz }}\d^2k\, \mathcal{F}^+(\tilde\bk)\,.
\end{equation}
Here, $\mathcal{F}^+$ is the sum of the Berry curvatures of all occupied bands with mirror eigenvalue $+\i$,
\begin{equation}
\mathcal{F}^+(\tilde\bk)=\sum_a\mathcal{F}^+_{a}(\tilde\bk)\,,\label{eq:berry}
\end{equation}
and $\overline{\rm \bz }$ is the surface Brillouin zone (\sbz) for the plane invariant under the mirror operation. Because $C^-=-C^+$ due to time-reversal symmetry, we could have also chosen $C^-$ to define the mirror Chern number.

\medskip

For cubic symmetry, there are three independent mirror planes, $z=0$, $z=1/2$, and $x=y$, leaving invariant the planes $k_z=0$, $k_z=\pi$, and $k_x=k_y$, respectively. For our definition of the spinor, the mirror operators are given by
\begin{subequations}
\begin{align}
M_z&=\i\,\tau_z\sigma_z\,,\\
M_{x-y}&=\i\,\tau_z\,\frac{\sigma_x-\sigma_y}{\sqrt2}\,.
\end{align}
\end{subequations}
For each of the different planes, there exists an associated mirror Chern number. We note that for planes with ${\rm C}_{n}$ symmetry, the (mirror) Chern number can be calculated up to a multiple of $n$ by multiplying rotational eigenvalues at the $n$-fold rotation-invariant points in the \sbz.\cite{fang_bulk_2012,ye_tci_2013} Here, we instead compute $C^+$ exactly for our model by using a numerical method for a discretized \bz.\cite{suzuki_chern_2005} The results are shown in \tab{tab:invariants} for all different phases.

The (momentum-resolved) Berry curvature of the bands with mirror eigenvalue $+\i$ is shown in \fig{fig:berry}(b) for the three different planes $k_z=0$, $k_z=\pi$, and $k_x=k_y$ and three different sets of parameters specified in \fig{fig:berry}(a). When the energy gap closes and reopens at one of the \hsp s, thereby creating an additional band inversion at that point, the Berry flux changes by $\pm 1$ at the respective point which leads to a change of the mirror Chern number by $\pm 1$.
We can therefore view the inverted \hsp s as sources of the Berry flux (monopoles). 

It is then possible to formulate a simple rule, illustrated in \fig{fig:charges}, which provides the three mirror Chern numbers for all eight different phases in \tab{tab:invariants}. Every mirror Chern number can be obtained by summing up all the ``charges" $\pm1$ at the inverted \hsp s, which lie in the respective mirror plane. As the relative sign of the charges at the different \hsp s is fixed by the condition that the trivial band insulator must have vanishing mirror Chern numbers, this rule is universal for a two-band model in the presence of cubic symmetry. The picture shown in \fig{fig:charges} is equivalent to the following formulas in terms of the band-inversions at the \hsp s for the mirror Chern numbers:
\begin{subequations}
\begin{align}
C_{k_z=0}^+&=w(\pt\Gamma)-2w(\pt X)+w(\pt M)\,,\\
C_{k_z=\pi}^+&=w(\pt X)-2w(\pt M)+w(\pt R)\,,\\
C_{k_x=k_y}^+&=w(\pt\Gamma)-w(\pt X)-w(\pt M)+w(\pt R)\,.
\end{align}\label{eq:mirror-rules}
\end{subequations}

We observe that the knowledge of the two mirror Chern numbers for $k_z=0$ and $k_z=\pi$ suffices to uniquely identify the topological phase and the parities of occupied bands at all \hsp s. In particular, also the $\zz$-invariants [\eq{eq:Z2}] follow:
\begin{subequations}
\begin{align}
(-1)^{\nu_0}&=(-1)^{C^+_{k_z=0}+C^+_{k_z=\pi}}\,,\\
(-1)^{\nu_{1,2,3}}&=(-1)^{C^+_{k_z=\pi}}\,.
\end{align}
\end{subequations}
Equations~\eqref{eq:mirror-rules} also directly provide the sum rule $C^+_{k_x=k_y}=C^+_{k_z=0}+C^+_{k_z=\pi}$ for the three mirror Chern numbers.

Note that in contrast to the $\zz$ indices, the mirror Chern numbers depend on the parity of the occupied bands; they are multiplied by $-1$ when switching $d$  and $f$ electrons. The sign of the mirror Chern number, also known as the \emph{mirror chirality}, is related to the direction of propagation of the surface states.\cite{teo_surface_2008,hsieh_majorana_2012} This effect will be further discussed in the following section.

\subsection{Surface states}\label{sec:surface}

\newsavebox{\figa}
\newsavebox{\figb}
\savebox{\figa}{%
\begin{tikzpicture}[scale=.8]
\draw (-2,-2) -- (-2,2) -- (2,2) -- (2,-2) -- cycle;
\draw[->] (-2.4,0) -- (2.4,0) node[right=.2cm]{$k_y$}; \draw[->] (0,-2.4) -- (0,2.4) node[above=.1]{$k_z$};
\draw[blue,dashed,thick] (-2,2) -- (2,2) (-2,0) -- (2,0) (2,-2) -- (2,2) (0,-2) -- (0,2) (-2,-2) -- (2,2);
\draw[text=blue,font=\scriptsize] (2,.8) node[above,rotate=90]{$k_y=\pi$} (1,2) node[above]{$k_z=\pi$} (0,1.2) node[above,rotate=90] {$k_y=0$} (1,0) node[below] {$k_z=0$} (1,1) node[above, rotate=45] {$k_y=k_z$};
\draw[text=red,font=\scriptsize] (0,0) node[above left] {$\bf \Gamma X$} (2,0) node[above right] {$\bf XM$} (0,2) node[above left] {$\bf XM$} (2,2) node[above right] {$\bf MR$};
\filldraw (0,0) circle (.04cm) node[below right]{\ptt{\bar\Gamma}} (2,0) circle (.04cm) node[below right]{\ptt{\bar X}} (0,2) circle (.04cm) node[below right]{\ptt{\bar X}} (2,2) circle (.04cm) node[below right]{\ptt{\bar M}};
\end{tikzpicture}}
\savebox{\figb}{%
\begin{tikzpicture}[scale=.8,x=0.707cm]
\draw (-2,-2) -- (-2,2) -- (2,2) -- (2,-2) -- cycle;
\draw[->] (-2,0)++(-.4cm,0) -- (2,0) -- ++(.4cm,0) node[right=.1cm]{$k_{\tilde y}$}; \draw[->] (0,-2.4) -- (0,2.4) node[above=.1]{$k_z$};
\draw[blue,dashed,thick] (-2,2) -- (2,2) (-2,0) -- (2,0) (0,-2) -- (0,2);
\draw[text=blue,font=\scriptsize] (1,2) node[above]{$k_z=\pi$} (0,1.2) node[above,rotate=90] {$k_y=k_x$} (1,0) node[above] {$k_z=0$};
\draw[text=red,font=\scriptsize] (0,0) node[above left] {$\bf \Gamma M$} (2,0) node[above right] {$\bf X^2$} (0,2) node[above left] {$\bf XR$} (2,2) node[above right] {$\bf M^2$};
\filldraw (0,0) circle (.04cm) node[below right]{\ptt{\bar\Gamma}} (2,0) circle (.04cm) node[below right]{\ptt{\bar X}} (0,2) circle (.04cm) node[below right]{\ptt{\bar Y}} (2,2) circle (.04cm) node[below right]{\ptt{\bar S}};
\end{tikzpicture}}

\begin{figure}[t]
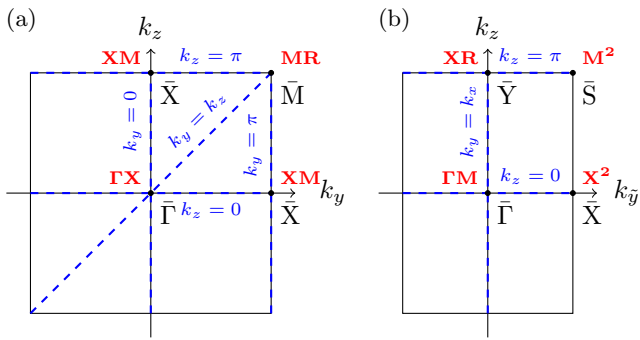

\centering
\figlabel{\usebox{\figa}}{a}{0,0}\hspace{2mm}
\figlabel{\usebox{\figb}}{b}{0,0}
\caption{(Color online) Surface \bz\ for the (100) (a) and (110) surfaces (b) of a simple cubic lattice. The dashed blue lines are orthogonal projections of mirror planes. The bulk \hsp s which are projected to each surface \hsp\ are displayed in red.}
\label{fig:surfaceBZ}
\end{figure}

The $\zz$ invariants $(\nu_0;\nu_1,\nu_2,\nu_3)$, as well as the mirror Chern numbers, imply gapless boundary states on all or only certain high-symmetry surfaces of the clean system. For example, a Dirac cone is present at all (high-symmetry) points in the \sbz\ onto which an odd number of bulk inverted \hsp s is projected.\cite{fu_topological_2007,jiang_observation_2013} On the other hand, a nonzero mirror Chern number implies that at least $|C^+|$ Dirac cones exist along the high-symmetry line (\hsl) in the \sbz , which is invariant under the respective mirror operation.\cite{hsieh_tci_2012} Strictly speaking, in the presence of disorder, only the strong invariant $\nu_0$ is well-defined, since its definition only requires time-reversal symmetry and charge conservation. However, it has been shown that both the weak $\zz$ invariants as well as the mirror Chern numbers retain their meaning if the disorder is sufficiently small and preserves the translation and/or mirror symmetries on average.\cite{ringel_strong_2012,fulga_statistical_2012} This robustness against weak disorder is also consistent with the recent experimental observations of an even number of Dirac cones on surfaces of SnTe, Pb$_{1-x}$Sn$_{x}$Se, and Pb$_{1-x}$Sn$_{x}$Te, where disorder is certainly present.\cite{tanaka_experimental_2012,dziawa_topological_2012,xu_observation_2012}

\begin{figure}[t]
\centering
\figlabel{\includegraphics[height=.22\textwidth]{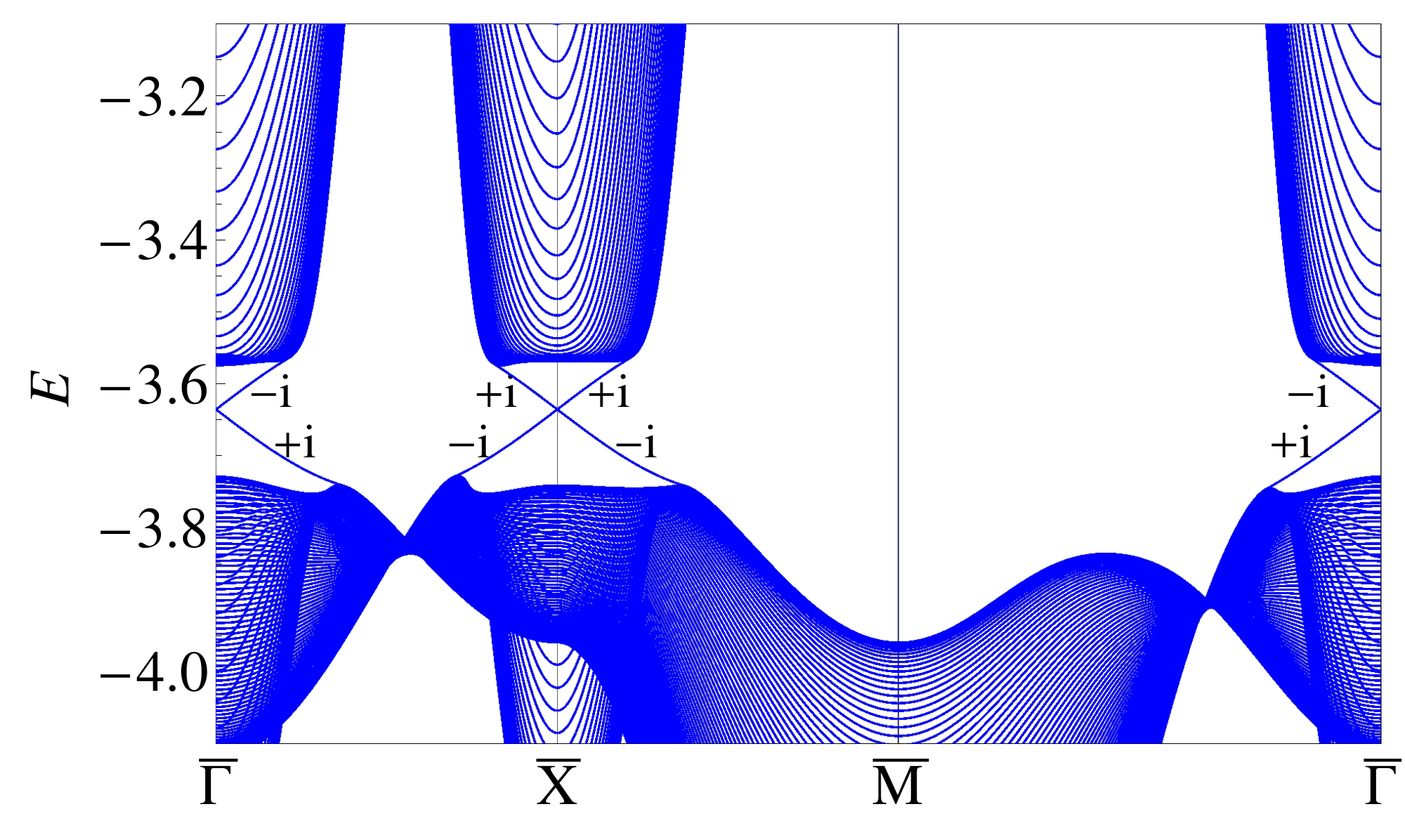}}{a}{0,0} \\[2mm]
\figlabel{\includegraphics[height=.22\textwidth]{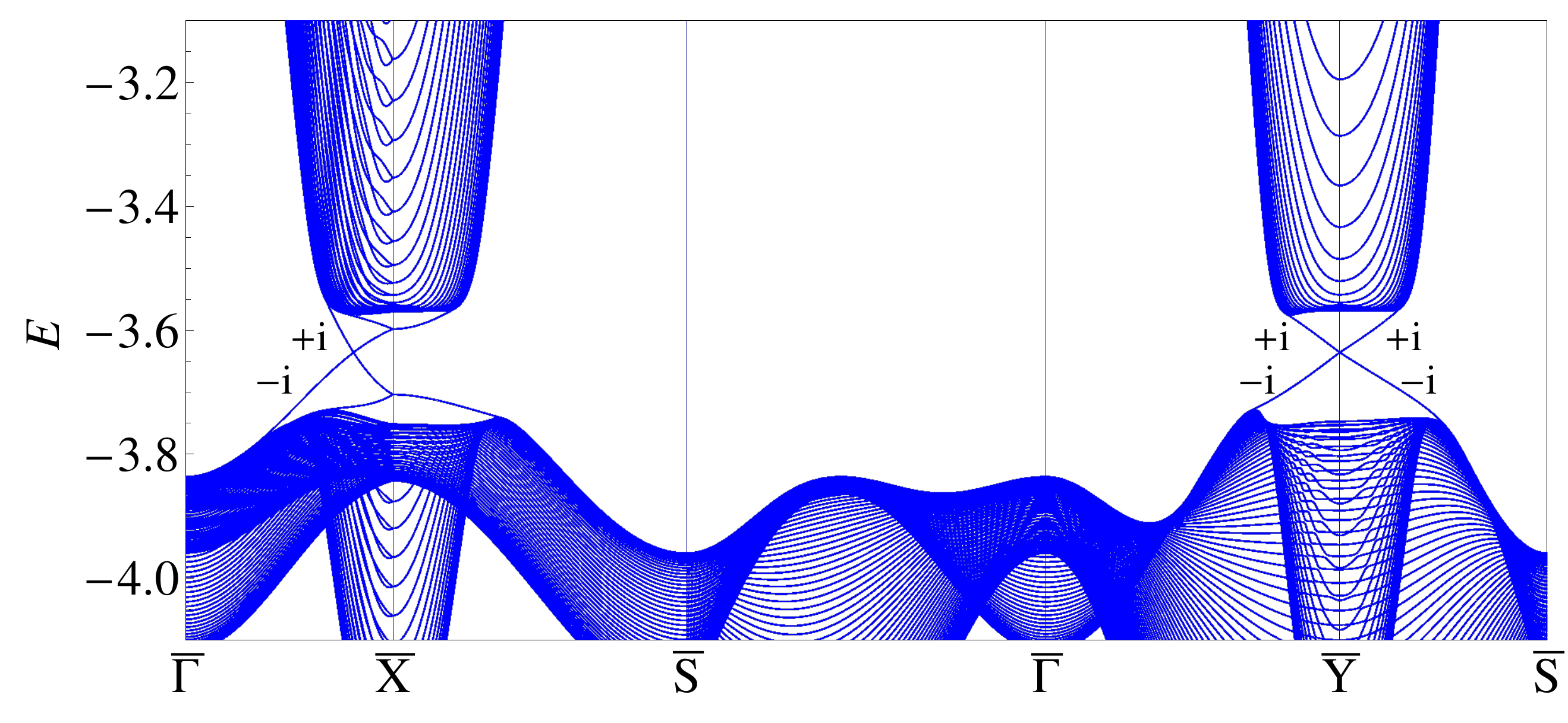}}{b}{0,0}
\caption{(Color online) Energy spectrum of a $100$ unit cells thick slab for the (100) (a) and (110) surface (b) in the phase \sti{X} ($\td=1$, $\tdd=-0.6$, $\tf=-0.1$, $\tff=0.06$, $V=0.1$, $\ef=-4$). For \hsl s that are projections of mirror planes, the surface states are labeled by the mirror eigenvalue of the state localized on the top (in the $+x$-direction) surface.
Note that the mirror operation differs for the various \hsl s such that the mirror eigenvalues do not have to coincide. All gapless modes predicted by the topological invariants are present. In the plot for the (110) plane, there is clearly visible the Dirac cone at the \ptt{\bar\Gamma}-\ptt{\bar X} line, which is protected by mirror symmetry.}
\label{fig:surface1}
\end{figure}

\begin{figure}[t]
\centering
\figlabel{\includegraphics[height=.22\textwidth]{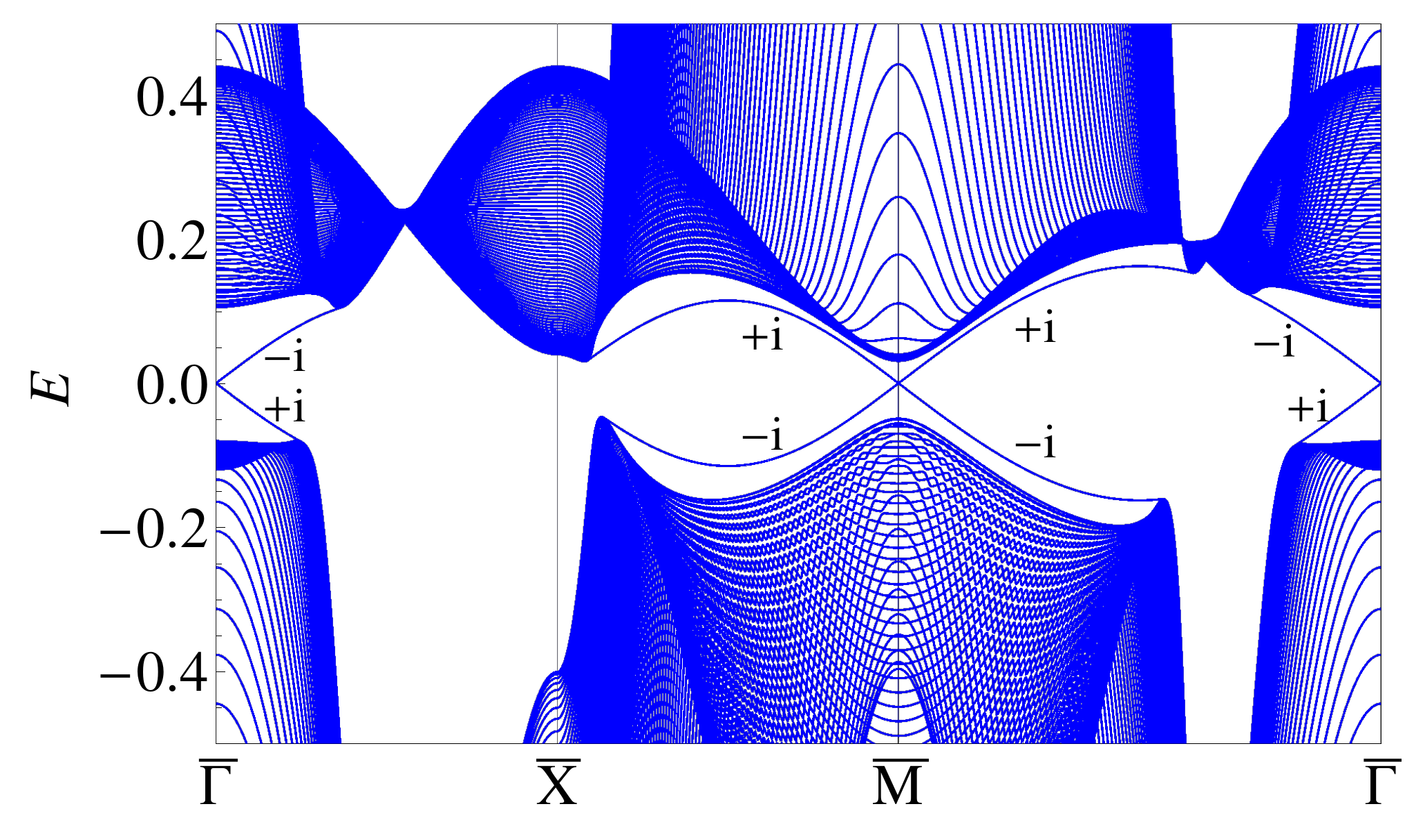}}{a}{0,0} \\[2mm]
\figlabel{\includegraphics[height=.22\textwidth]{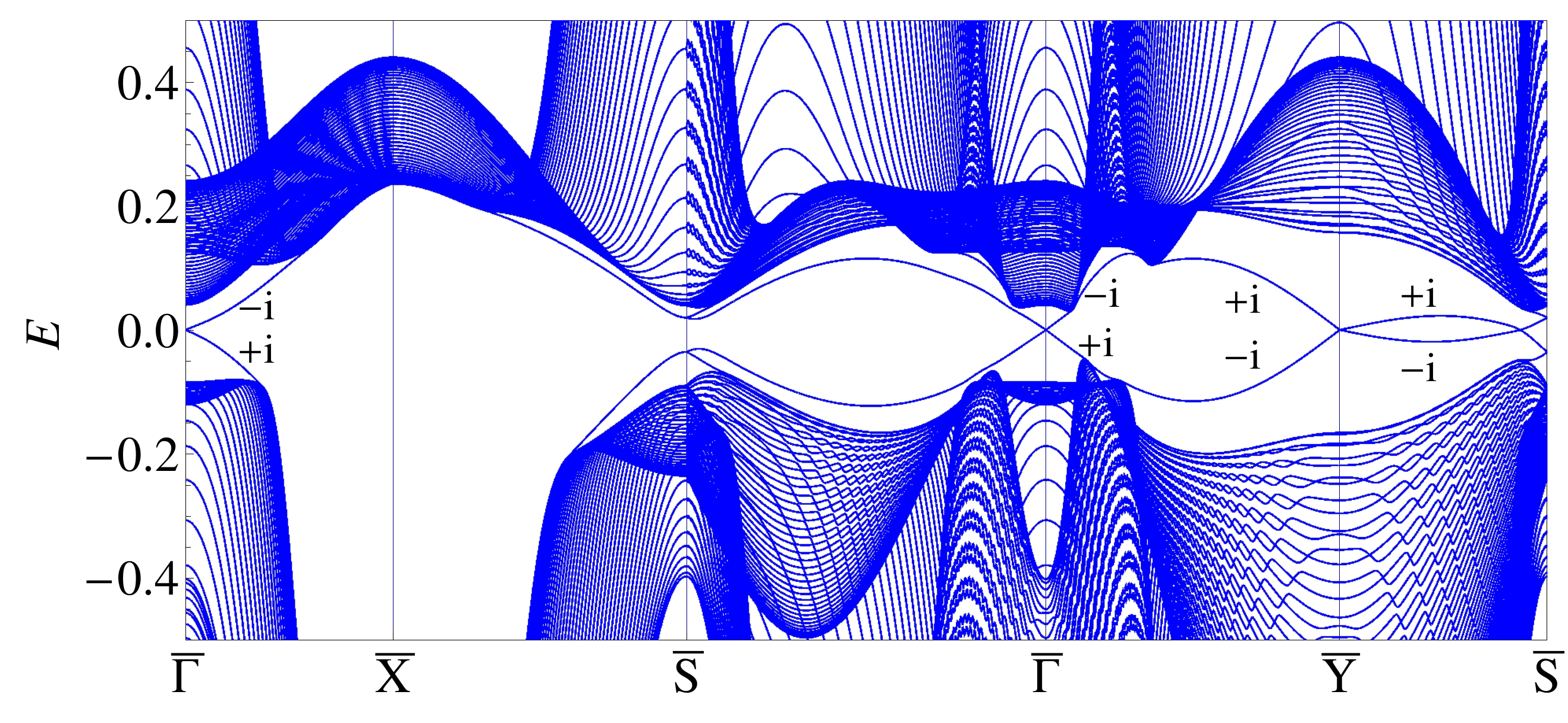}}{b}{0,0}
\caption{(Color online) Energy spectrum of a $100$ unit cells thick slab for the (100) (a) and (110) surface (b) in the phase \wtii{\Gamma R} (hopping and hybridization as in \fig{fig:surface1}, $\ef=0$). The states are labeled by $\pm\i$ using the same convention as in \fig{fig:surface1}. All gapless modes predicted by the topological invariants are present. On the (110) surface there is an additional crossing on the \ptt{\bar Y}-\ptt{\bar S} line that occurs due to the mirror chirality associated with the $k_z=\pi$ plane.}
\label{fig:surface2}
\end{figure}

\begin{figure}[t]
\centering
\figlabel{\includegraphics[height=.22\textwidth]{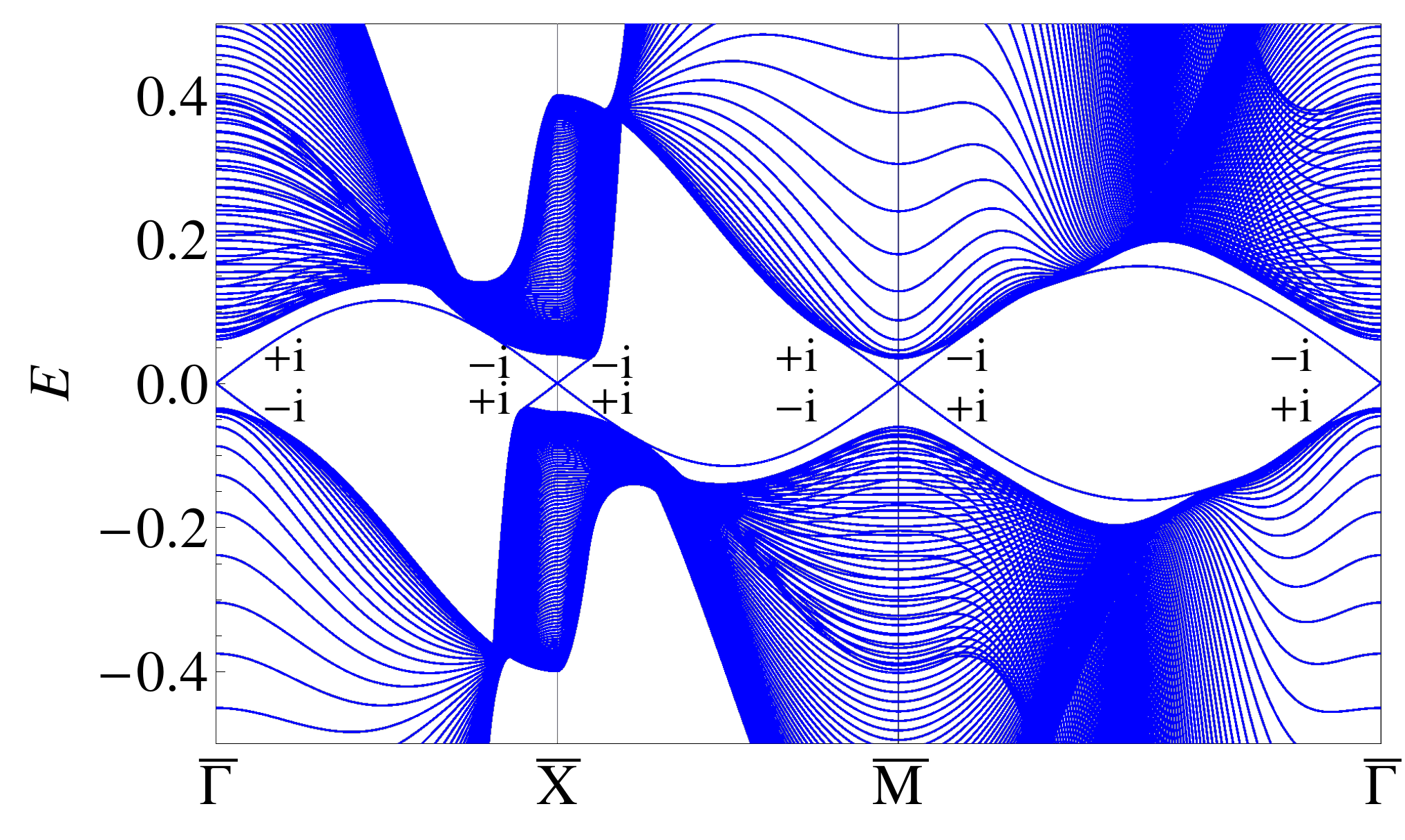}}{a}{0,0} \\[2mm]
\figlabel{\includegraphics[height=.22\textwidth]{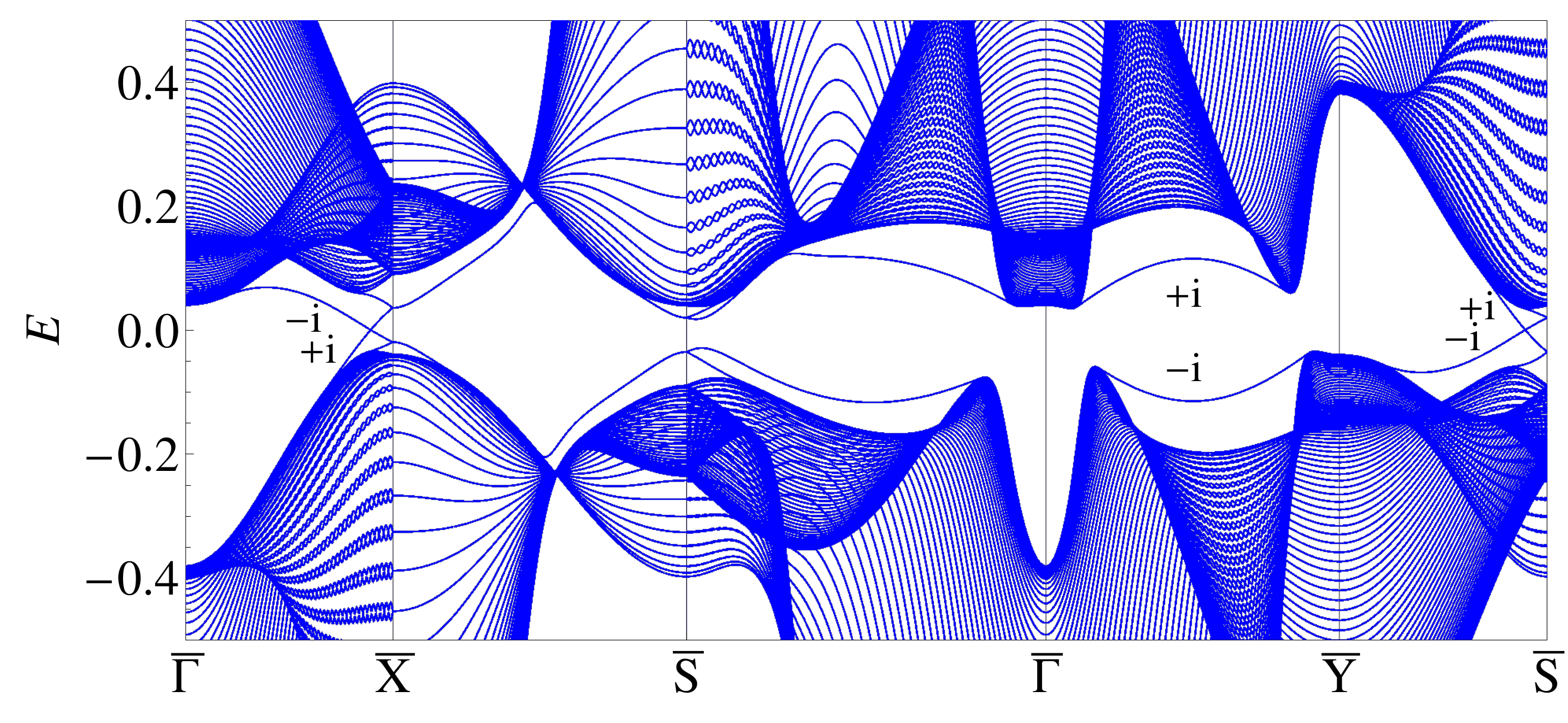}}{b}{0,0}
\caption{(Color online) Energy spectrum of a $100$ unit cells thick slab for the (100) (a) and (110) surface (b) in the phase \tci{\Gamma M} ($\td=1$, $\tf=-0.1$, $\ttt=0$, $\tttt/\tt=0.3$, $V=0.1$, $\ef=0$). The states are labeled by $\pm\i$ using the same convention as in \fig{fig:surface1}. There are four Dirac cones on each of the surfaces, which are all protected by mirror symmetry only (note that only two are visible on the (110) surface as only half of each \hsl\ is plotted).}
\label{fig:surface3}
\end{figure}

The sign of the mirror Chern number is called the mirror chirality.\cite{teo_surface_2008,hsieh_majorana_2012} It determines the mirror eigenvalue ($\pm \i$) of the surface bands crossing from the valence to the conduction band in the positive direction of the respective \hsl. If, e.g., the mirror Chern number is $C^+=+2$, two bands with eigenvalue $+\i$ cross from valence to conduction band. The positive direction is defined by the orientation of the mirror-invariant plane in the calculation of the mirror Chern number and the chosen surface as follows. First, in order to calculate the mirror Chern number, a right-handed coordinate system with a unit vector $\bs n_{\rm mp}$ perpendicular to the mirror-invariant plane needs to be defined. Second, for the surface, we define $\bs n_{\rm sf}$ as the outward pointing normal vector. Then, the positive direction of the \hsl s is defined by the vector-product $\bs n_{\rm sf}\times\bs n_{\rm mp}$. For our choice of coordinates, the positive direction is always from left, bottom to right, top in the drawings of \fig{fig:surfaceBZ}.

Figure~\ref{fig:surfaceBZ} shows the \sbz\ for the (100) and (110) surface along with the projected bulk \hsp s and projected mirror planes. Consulting \tab{tab:invariants}, it is now simple to determine if and where gapless Dirac points are expected in the \sbz. For the (100) surface, the location of the Dirac points predicted by the $\zz$-invariants agrees with the values of the three mirror Chern numbers. Thus, all the Dirac points are located at \hsp s of the (100) \sbz. On the other hand, for the (110) surface and if there is a band inversion at \ptt{X} (\ptt{M}), the $\zz$-invariants predict one Dirac cone located at \ptt{\bar Y} (\ptt{\bar \Gamma}). However, due to the nonzero mirror Chern numbers $|C_{k_z=0}|=2$ ($|C_{k_z=\pi}|=2$), there must be two additional Dirac cones on the $k_z=0$ ($k_z=\pi$) \hsl, which are protected by mirror symmetry.\cite{ye_tci_2013}

\medskip

We have confirmed these expectations by explicitly calculating the surface states in thin film geometries (see \app{sec:surface-calc} for further details). The results of the calculations for three different topological phases are shown in \fig{fig:surface1}, \fig{fig:surface2}, and \fig{fig:surface3}. 

Figure~\ref{fig:surface1} shows the energy spectrum for (a) the (100) and (b) the (110) surface of the strong topological insulator \sti{X}. As expected, there are Dirac cones at \ptt{\bar\Gamma} and the two \ptt{\bar X} points for the (100) surface. On the other hand, for the (110) surface, we observe additional two Dirac cones on the \ptt{\bar X}-\ptt{\bar\Gamma}-\ptt{\bar X} line protected by the mirror symmetry.\cite{ye_tci_2013} (Note that only one Dirac cone is visible because only half of the \hsl\ corresponding to the $k_z=0$ plane is shown.)

We now turn to the surface spectrum of the weak topological insulator \wtii{\Gamma R}, shown in \fig{fig:surface2}, which has a remarkable feature on its (110) surface. For this phase, the $\zz$-invariants predict Dirac cones at the \ptt{\bar\Gamma} and \ptt{\bar X} points. These Dirac cones are also required by the mirror Chern numbers. Nevertheless, we find two additional (and unexpected) band crossings along the \ptt{\bar S}-\ptt{\bar Y}-\ptt{\bar S} line. The solution to this puzzle lies in the sign of the mirror Chern number, the mirror chirality. The mirror Chern number for the $k_z=\pi$ plane is $-1$, which means that along the line \ptt{\bar S}-\ptt{\bar Y}-\ptt{\bar S} one band with mirror eigenvalue $+\i$ crosses from the conduction to the valence band. However, as shown in the \app{sec:kp} within the ${\bs k}\cdot {\bs p}$ theory, the velocity of this band at the \ptt{\bar Y} point is given by 
\begin{equation}
v=\frac{4|V|\sqrt{-\tf\td}}{\td-\tf}\,,
\label{eq:v}
\end{equation}
which is positive for $\td>0$ and $\tf<0$. Therefore, the two bands with eigenvalues $\pm\i$ crossing at the \ptt{\bar Y} point must cross \emph{two} additional times along the \ptt{\bar S}-\ptt{\bar Y}-\ptt{\bar S} line in order to fulfill the constraint from the mirror chirality. We note that a similar observation was made for the surface states of a time-reversal-invariant topological superconductor.\cite{hsieh_majorana_2012}

Finally, \fig{fig:surface3} shows the surface states of the topological crystalline insulator phase \tci{\Gamma M}. In this phase, all $\zz$ invariants vanish. Therefore, all gapless surface states are protected by mirror symmetry. Due to the mirror Chern numbers $C^+_{k_z=0}=2$ and $C^+_{k_z=\pi}=-2$, there are two Dirac cones on each of the corresponding \hsl s.

\section{Interaction effects}\label{sec:interactions}
\subsection{Renormalization of band parameters}\label{sec:renorm}
In the previous section, we have analyzed the topological properties of the noninteracting Hamiltonian, keeping in mind that the band parameters are potentially renormalized by the interactions among the $f$ electrons. In the following, we want to explicitly study {\it how} the renormalization parameters depend on the interaction $U$ and the noninteracting band parameters. For this purpose, we now consider again the full Hamiltonian~\eqref{eq:model_general} including the interaction part $H_{\rm int}$ given in \eq{eq:Hint}.
We treat the resulting problem, which is now quartic in creation and annihilation operators, within the Kotliar-Ruckenstein slave-boson scheme.\cite{kotliar_new_1986} Details of this method are presented in \app{sec:kotliar-ruckenstein}, but in essence, it leads to a new (noninteracting) mean-field Hamiltonian with the renormalized parameters $\tilde{\tf}=z^2\tf$, $\tilde{\tff}=z^2\tff$, $\tilde{\tfff}=z^2\tfff$, $\tilde{V}=zV$, and $\tilde{\ef}=\ef+\lambda$, and a total energy offset of $N(\nd U-\lambda \nf)$:
\begin{equation}
H\rightarrow H_{\rm ni}(\tilde{\tf},\tilde{V},\tilde{\ef})+N(\nd U-\lambda \nf)\,.\label{eq:model_renorm}
\end{equation}
Here, the renormalization factor $z$ depends on the double occupancy $\nd$ and the filling of the $f$ orbitals $\nf$ and assumes the form known from the Gutzwiller approximation:\cite{Rice:1985b}
\begin{equation}
z=\frac{\sqrt{\nd\left(\frac{\nf}{2}-\nd\right)}+\sqrt{(1-\nf+\nd)\left(\frac{\nf}{2}-\nd\right)}}{\sqrt{\frac{\nf}{2}\left(1-\frac{\nf}{2}\right)}}\,.
\label{eq:renorm_z}
\end{equation}
Because $0\leq z\leq1$, the renormalization factor $z$ reduces the hopping of $f$ electrons and the hybridization. From Eqs.~\eqref{eq:model_renorm} and \eqref{eq:renorm_z}, the shift $\lambda$, the double occupancy $\nd$ and the occupation of $f$ orbitals $\nf$ are obtained self-consistently as a function of the interaction $U$ and the noninteracting band parameters from the saddle-point approximation for the ground-state energy (see \app{sec:kotliar-ruckenstein}).
\begin{figure}                  
\centering
\figlabel{\includegraphics[width=.45\textwidth]{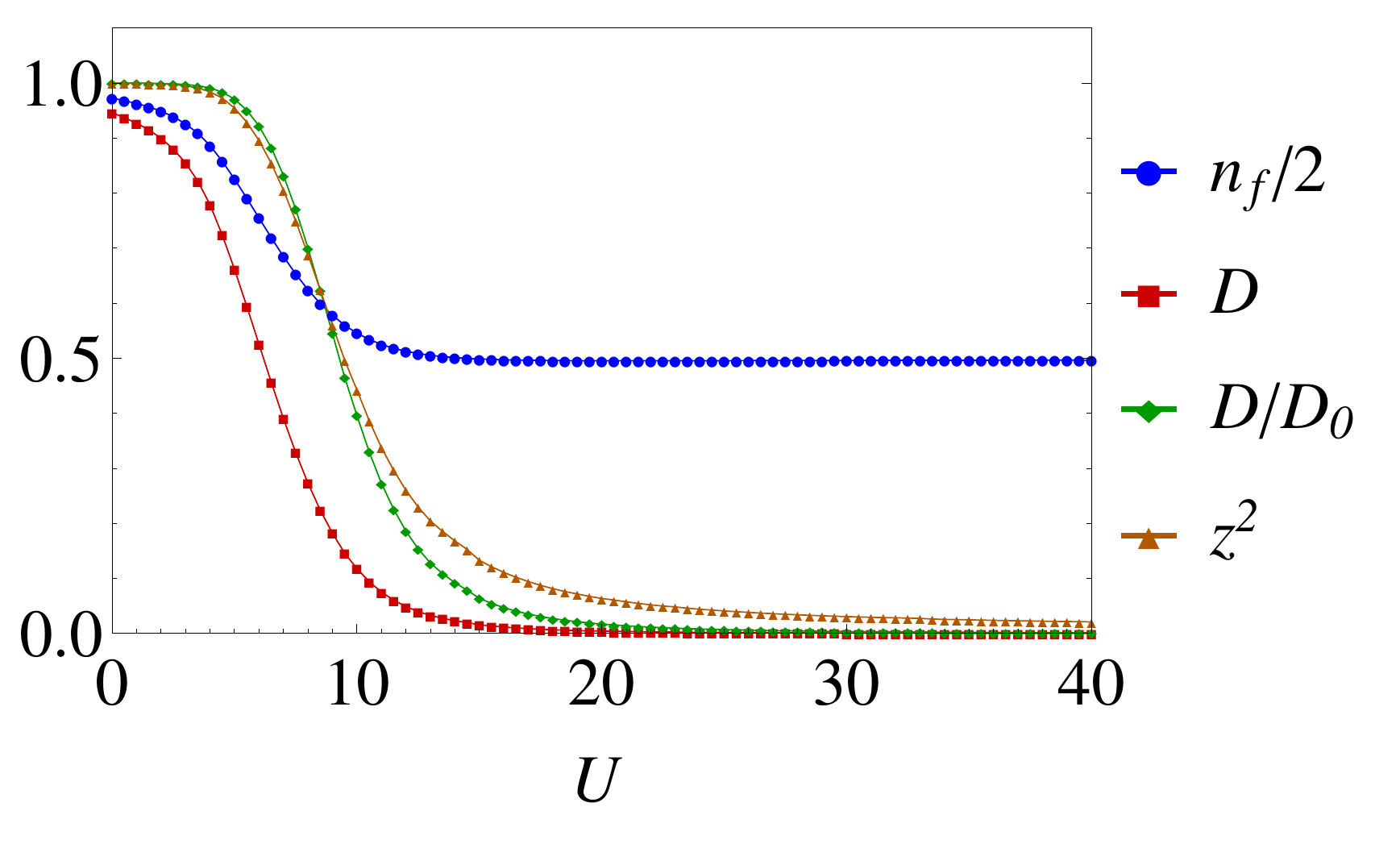}}{a}{-3mm,0}\\[2mm]
\figlabel{\includegraphics[width=.45\textwidth]{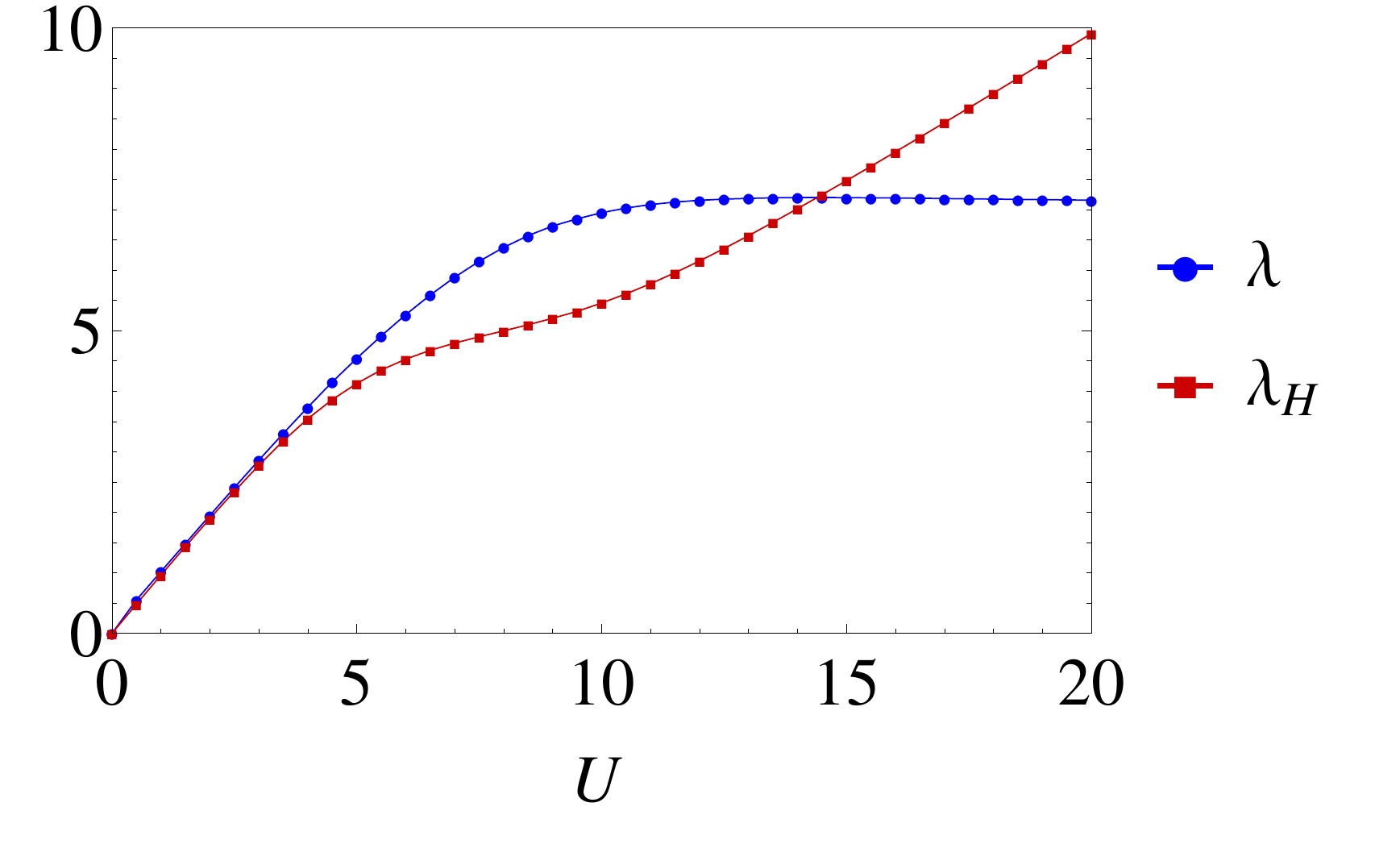}}{b}{-3mm,0}
\caption{(Color online) Plot of occupation numbers and hopping renormalization (a) and comparison of the renormalization of the chemical potential and the Hartree contribution (b) for $\td=1$, $\tdd=-0.4$, $\tf=-0.1$, $\tff=0.04$, $\tttt=0$, $\ef=-8$, and $V=0.5$. For small interaction there is almost perfect agreement of $\lambda$ with $\lambda_{\rm H}$, while for large $U$ the renormalization $\lambda$ converges against some finite value. }
\label{fig:plotu}
\end{figure}

\begin{figure}
\centering
\includegraphics[width=.45\textwidth]{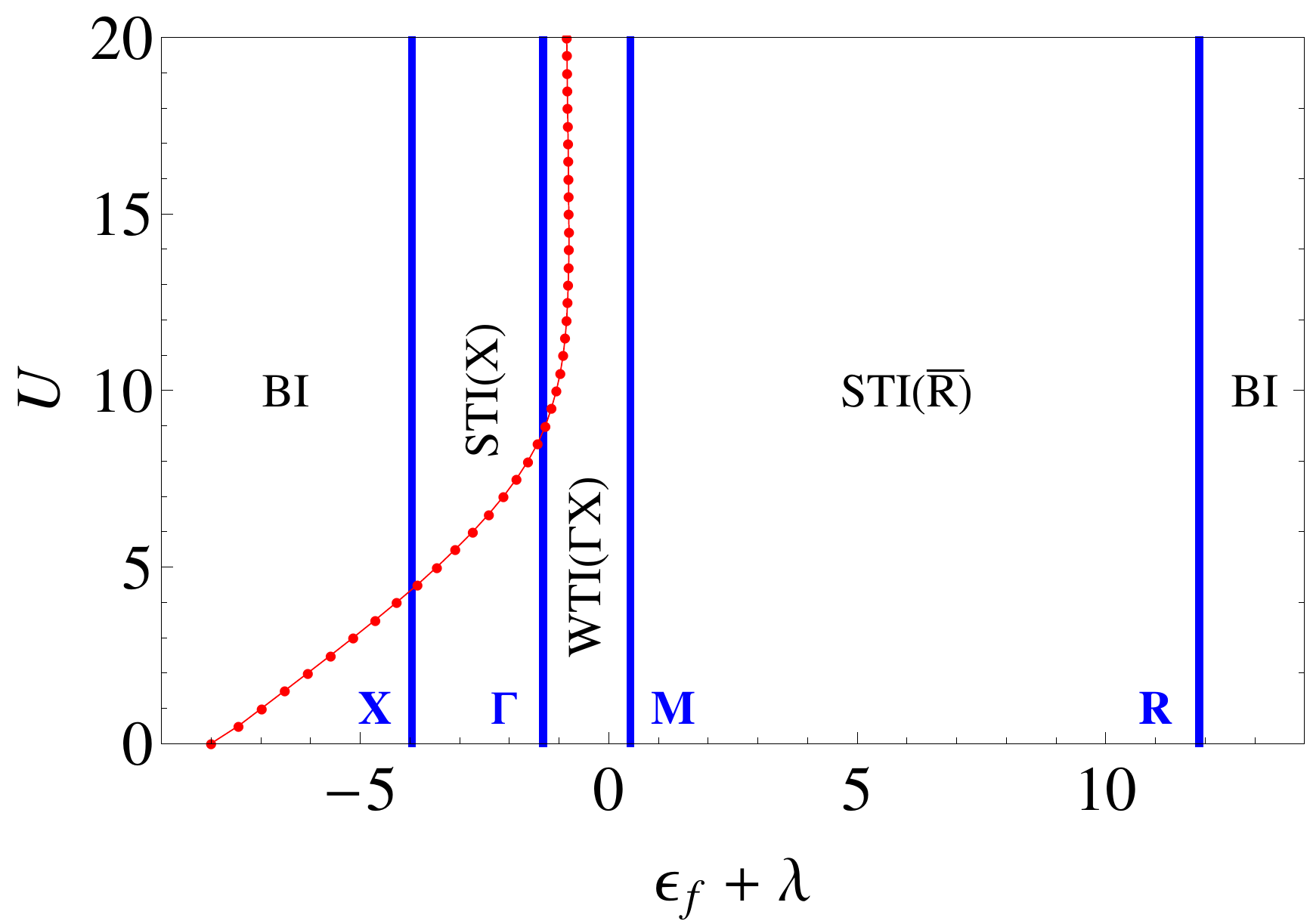}
\caption{(Color online) The renormalized chemical potential $\ef+\lambda$ is shown in red (light gray) for different values of the interaction $U$. The noninteracting band parameters were chosen as in \fig{fig:plotu} and remain fixed. At $U=0$ the system is in the trivial phase \bi, for different finite values of $U$ it undergoes topological phase transitions to the phases \sti{X} and \wti{\Gamma X}.}
\label{fig:phasetransu2}
\end{figure}

In \fig{fig:plotu}(a), we show the self-consistent values of $\nd$, $\nf$, and $z$ as a function of the interaction $U$ for noninteracting band parameters in the \bi\ phase. For repulsive interaction, $U>0$, the double occupancy $\nd$ is reduced below its noninteracting value
\begin{equation}
\nd_0=\langle\hat n_{i\uparrow}\rangle\langle\hat n_{i\downarrow}\rangle=\frac{\nf^2}{4},\ 
\end{equation}
and hence $\nd/\nd_0<1$. This suppression of $\nd$ leads to a reduction of the renormalization factor, $z<1$. As a consequence, the (weak) dispersion of the $f$ electrons is further suppressed and the hybridization gap is reduced by the interactions. In \fig{fig:plotu}(b), we show the shift $\lambda$ of the $f$-orbital level. For small $U$, this shift follows the Hartree contribution $\lambda_{\rm H}=U\nf/2$, but clearly deviates for stronger interactions where it saturates. This dependence of $\lambda$ on $U$ is also reflected in the  dependence of $\nf$ on $U$ shown in \fig{fig:plotu}(a): $\nf$ is reduced for small interactions but then approaches a constant value. 

\subsection{Interaction-driven topological phase transitions}

\begin{figure}
\centering
\includegraphics[width=.45\textwidth]{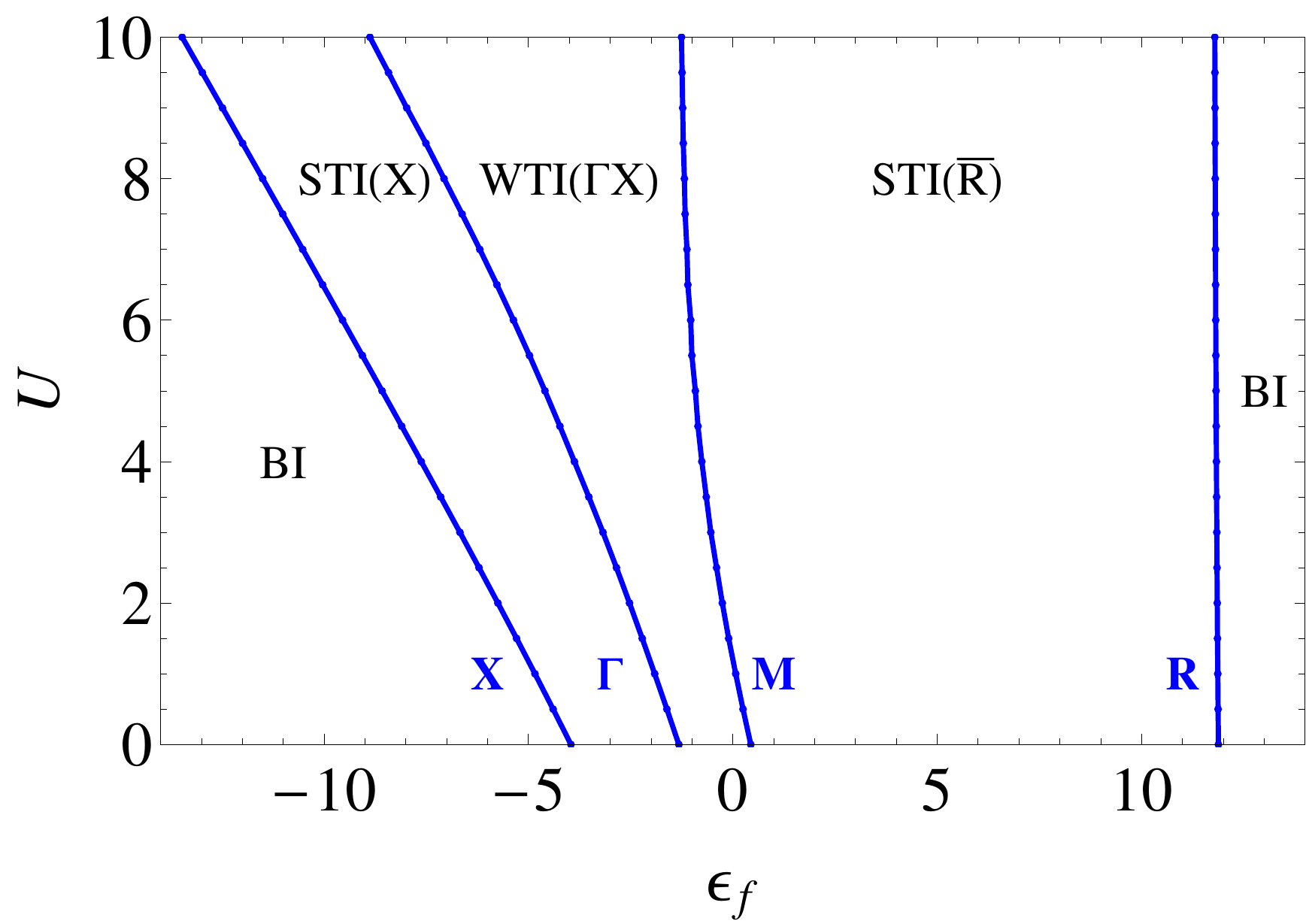}
\caption{(Color online) Phase transitions of the full Hamiltonian~\eqref{eq:model_general} in a mean-field treatment of the slave bosons for $\td=1$, $\tdd=-0.4$, $\tf=-0.1$, $\tff=0.04$, $\tttt=0$, and $V=0.5$. 
The labeling of phases and phase transitions follows the same convention as in \fig{fig:phases_tf_ef}.
The occupation of $f$ orbitals is virtually constant along the lines of phase transitions as the renormalized chemical potential for the $f$ electrons remains constant along those lines (see also \tab{tab:phasetrans_nf}).}
\label{fig:phasetransu}
\end{figure}

The shift of the $f$-electron level by $\lambda$ with $U$ can crucially affect the topological properties of the renormalized quasiparticle Hamiltonian, because the points with band inversion may change. In fact, the interaction may drive topological phase transitions between various trivial and topological phases. Figure~\ref{fig:phasetransu2} shows the topological phase transitions for one particular choice of noninteracting band parameters.

In \fig{fig:phasetransu}, we show the $\ef$-$U$ phase diagram for a fixed choice of noninteracting band parameters, illustrating how the lines of phase transitions bend with increasing $U$. We note that in this context, the renormalization of the hopping of $f$ electrons by a factor $z^2$ and the hybridization by a factor of $z$ does not affect the topology of the system. Similar interaction-driven phase transitions have also been observed in interacting versions of quantum spin Hall models\cite{wang_interaction_2012,budich_fluctuation_2012,budich_fluctuation_2013} and a two-dimensional model of a topological Kondo insulator.\cite{werner_interaction_2013,werner_temperature_2013}

\begin{table}
\caption{Comparison of occupation of $f$ orbitals for different values of the interaction for the four lines of phase transitions shown in \fig{fig:phasetransu}. The occupations change very little when moving along a line of phase transition for varying $U$. Therefore, the filling $n$ is a good indicator of the phase for fixed hopping and hybridization.}
\begin{ruledtabular}
\begin{tabular}{cccccc}
&&\ptt{X}&\ptt{\Gamma}&\ptt{M}&\ptt{R}\\\hline
\multirow{2}{*}{$U=0$}&$\ef$&-3.96&-1.32&0.44&11.88\\&$\nf/2$&0.865&0.578&0.387&0.016\\\hline
\multirow{2}{*}{$U=10$}&$\ef$&-12.98&-8.42&-1.24&11.79\\&$\nf/2$&0.836&0.559&0.375&0.016
\end{tabular}
\end{ruledtabular}
\label{tab:phasetrans_nf}
\end{table}

The results in \fig{fig:phasetransu} show that for a fixed onsite potential $\ef$, the topological phase also depends on the interaction strength $U$. Therefore, the value of $\ef$ does not provide a robust indicator of the topological phase. Remarkably, however, along the lines of phase transitions in the $\ef$-$U$ diagram, the filling of $f$ orbitals stays almost constant (see \tab{tab:phasetrans_nf}). This is due to the fact, that it much more strongly depends on the renormalization of the chemical potential than on the renormalization of the hopping and hybridization parameters. Therefore, for fixed noninteracting hopping and hybridization parameters, the filling $\nf$ is a more useful indicator of the topological phase than $\ef$.

We also remark that when starting from fixed noninteracting band parameters in the mixed-valence regime with $\nf>1$, one generically ends up in the local-moment regime at $\nf\approx 1$ for large interactions. This is because the electron-electron interaction among the $f$ electrons suppresses double occupancy but also pushes electrons from the correlated $f$ band into the dispersive $d$ band by renormalizing the $f$-orbital level (see Fig.~\ref{fig:plotu}). As discussed in \Sec{sec:phase-diagrams} and shown in \fig{fig:phases_tff_ef}(b), the weak topological insulator phases are strongly favored if $n_f\approx 1$. As a consequence, a strong topological insulator is less likely to be realized in the large $U$ limit.

\section{Conclusion}\label{sec:conclusion}
In this paper, we constructed a minimal two-orbital model in the form of the periodic Anderson model for a cubic topological Kondo insulator. We found eight topologically distinct Kondo insulator phases, depending on the inversions of the two bands, which can be characterized by $\zz$ invariants (protected by time-reversal symmetry) and mirror Chern numbers (protected by different mirror symmetries). These topological invariants have a direct influence on the existence and position of gapless surface modes, which was demonstrated by numerically diagonalizing the Hamiltonian for a slab of finite thickness.

We studied the effect of interactions among the localized $f$ electrons using the Kotliar-Ruckenstein slave-boson scheme. We demonstrated how the renormalization of the chemical potential of $f$ electrons changes the band-inversions and thereby the topology of the system. Depending on the noninteracting band parameters, the interactions among electrons can destroy or facilitate a topological phase. As the occupation of $f$ orbitals is mainly determined by the renormalized chemical potential and only weakly depends on the hopping and hybridization amplitudes, it stays almost constant along the lines of phase transitions for variable $U$ and therefore provides a more robust criterion for the topological phase than, e.g., the bare chemical potential.

An interesting direction for future studies can be the extension to finite temperatures as the
gap of the Kondo insulator is of dynamical nature and will close eventually upon increasing 
temperature. Furthermore, following Doniach's picture of the competition between Kondo 
screening and \abb{rkky} interaction, a quantum phase transition between the Kondo insulating 
and a magnetically ordered phase can occur.\cite{doniach_kondo_1977} Topological Kondo insulators may show novel 
intriguing behavior at this transition. For the comparison with experiments of real materials such 
as \smb, our model needs to be extended to take a more close approach to the real electronic structure.\\\\

\section*{Acknowledgments}
We acknowledge a helpful discussion with A.\ Bernevig. This work was supported by a grant and the Ambizione program of the Swiss National Science Foundation.


\appendix
\section{Calculation of surface states}\label{sec:surface-calc}
We now want to explicitly calculate the surface energy spectrum by considering slabs which have periodic boundary conditions along two directions and are finite (with open boundary conditions) along the third. To simplify the formulas we define
\begin{subequations}
\begin{align}
h_0&=\ef\,\frac{\E-\tau_z}{2} \,,\\
h_{\rm 1r}&=-\td\,\frac{\E+\tau_z}{2}-\tf\,\frac{\E-\tau_z}{2} \,,\\
h_{\rm 1i}^\alpha&=V\tau_z\sigma_\alpha \,,\\
h_2&=-\tdd\,\frac{\E+\tau_z}{2}-\tff\,\frac{\E-\tau_z}{2} \,,\\
h_3&=-\tddd\,\frac{\E+\tau_z}{2}-\tfff\,\frac{\E-\tau_z}{2} \,.
\end{align}
\end{subequations}
Then, we can write the Hamiltonian~\eqref{eq:model_non-int} as
\begin{equation}
\begin{split}
H_{\rm ni}&=\sum_i\Psi_i^\dagger\,h_0\,\Psi_i+\Bigg[\sum_{\alpha=x,y,z}\sum_{\langle i,j\rangle_\alpha}\Psi_i^\dagger\left(h_{\rm 1r}+\i h_{\rm 1i}^\alpha\right)\Psi_j\\
&\quad +\sum_{\Nnn{i,j}}\Psi_i^\dagger\,h_2\,\Psi_j+\sum_{\Nnnn{i,j}}\Psi_i^\dagger\,h_3\,\Psi_j\ +\hc\Bigg]\,.\label{eq:model_simplified}
\end{split}
\end{equation}

For the $(100)$ surface, we can now do a Fourier transform of this Hamiltonian along the $y$ and $z$ directions. This leads to
\begin{subequations}
\begin{equation}
\begin{split}
&H_{\rm ni}=\sum_{x,k_y,k_z}\Big[\Psi^\dagger(x,k_y,k_z)\,h_0^{(100)}(k_y,k_z)\,\Psi(x,k_y,k_z)\\
 &\ +\!\left(\Psi^\dagger(x,k_y,k_z)\,h_1^{(100)}(k_y,k_z)\,\Psi(x\!+\!1,k_y,k_z)+\hc\right)\Big],
\end{split}
\end{equation}
where the matrices $h_0^{(100)}$ and $h_1^{(100)}$ are given by
\begin{align}
\begin{split}
h_0^{(100)}(k_y,k_z)&=h_0+2(c_y+c_z)h_{\rm 1r}\\
&\quad-2\left(s_yh_{\rm 1i}^y+s_zh_{\rm 1i}^z\right)+4c_{yz}h_2\,,
\end{split}\\
h_1^{(100)}(k_y,k_z)&=h_{\rm 1r}+\i h_{\rm 1i}^x+2(c_y+c_z)h_2+4c_{yz}h_3 .
\end{align}
\label{eq:hamiltonian100}
\end{subequations}

For the $(110)$ surface, we first have to define a new variable $\tilde{y}:=1/\sqrt{2}(y-x)$, which lies in the surface-plane. Now we can again do a Fourier transform of \eqref{eq:model_simplified} along the $\tilde y$ and $z$ directions. This leads to
\begin{subequations}
\begin{equation}
\begin{split}
&H_{\rm ni}=\sum_{x,k_{\tilde y},k_z}\Big[\Psi^\dagger(x,k_{\tilde y},k_z)\,h_0^{(110)}(k_{\tilde y},k_z)\,\Psi(x,k_{\tilde y},k_z)\\
&\ +\left(\Psi^\dagger(x,k_{\tilde y},k_z)\,h_1^{(110)}(k_{\tilde y},k_z)\,\Psi(x\!+\!1,k_{\tilde y},k_z)+\hc\right)\\
&\ +\left(\Psi^\dagger(x,k_{\tilde y},k_z)\,h_2^{(110)}(k_{\tilde y},k_z)\,\Psi(x\!+\!2,k_{\tilde y},k_z)+\hc\right)\Big]\,,
\end{split}
\end{equation}
where the matrices $h_0^{(110)}$, $h_1^{(110)}$, and $h_2^{(110)}$ are given by
\begin{align}
\begin{split}
h_0^{(110)}(k_{\tilde y},k_z)&=h_0+2c_zh_{\rm 1r}-2s_zh_{\rm 1i}^z\\
&\quad+2\cos\left(\sqrt{2}k_{\tilde y}\right)\left(h_2+2c_{z}h_3\right)\,,
\end{split}\\
\begin{split}
h_1^{(110)}(k_{\tilde y},k_z)&=h_{\rm 1r}+\i h_{\rm 1i}^x+\e^{\i\sqrt{2}k_{\tilde y}}\left(h_{\rm 1r}+\i h_{\rm 1i}^y\right)\\
&\quad+2c_z\left(1+\e^{\i\sqrt{2}k_{\tilde y}}\right)h_2 \,,
\end{split}\\
h_2^{(110)}(k_{\tilde y},k_z)&=\e^{\i\sqrt{2}k_{\tilde y}}\left(h_2+2c_{z}h_3\right)\,.
\end{align}
\label{eq:hamiltonian110}
\end{subequations}

Now, the Hamiltonians~\eqref{eq:hamiltonian100} and \eqref{eq:hamiltonian110} can be diagonalized numerically in order to obtain the surface energy spectrum.

\section{The \textit{k$\cdot$p} theory on the (110) surface}\label{sec:kp}
In this appendix we derive the result~\eqref{eq:v} for the velocity of the surface states crossing the \ptt{\bar Y} point on the (110) surface in the \wti{\bar\Gamma\bar R} phase. For this purpose, we derive the ${\bs k}\cdot {\bs p}$ model 
by expanding around one of the three, by symmetry related, \ptt{X} points. We choose \ptt{X} $=(0,0,\pi)$ which is projected onto the \ptt{\bar Y} point on the (110) surface. For notational simplicity, we assume that $\tddd=\tfff=0$ in the following. The gap closes at the \ptt{X} point if
\begin{equation}
\ef=\ef(\pt{X})\equiv-2(\td-2\tdd-\tf+2\tff)\,.
\end{equation}
Denoting $2M=\ef-\ef(\pt{X})$ and expanding around the \ptt{X} point yields the following matrix:
\begin{widetext}
\begin{align}
&H_{\pt{X}}(\bk)=\\
&\begin{pmatrix}
-M+\td k^2-2(\td+2\tdd)k_z^2 & 0 & 2Vk_z & -2V k_-\\
0&-M+\td k^2-2(\td+2\tdd)k_z^2 & -2Vk_+ & -2Vk_z\\
2Vk_z & -2Vk_- & M+\tf k^2-2(\tf+2\tff)k_z^2 & 0\\
-2Vk_+ & -2Vk_z & 0 &  M+\tf k^2-2(\tf+2\tff)k_z^2
\end{pmatrix}\,.\nonumber
\end{align}
\end{widetext}
Here, we have introduced $k_{\pm}=k_x\pm i k_y$. For the (110) surface, the mirror plane $k_z=\pi$ is projected onto the line \ptt{\bar{Y}}-\ptt{\bar{S}} along which we observe a second crossing in the \ptt{XM}-inverted phase. We therefore use the eigenbasis of the mirror operator
$M_z=i\sigma_z\tau_z$.
Furthermore, we introduce the following new coordinates:
\begin{subequations}
\begin{align}
u&=\frac{1}{\sqrt{2}}(-x+y)\,,\\
w&=\frac{1}{\sqrt{2}}(x+y)\,.
\end{align}
\end{subequations}
Or, in reciprocal space,
\begin{subequations}
\begin{align}
k_u&=\frac{1}{\sqrt{2}}(-k_x+k_y)\,,\\
k_w&=\frac{1}{\sqrt{2}}(k_x+k_y)\,.
\end{align}
\end{subequations}
Then, along the \ptt{\bar{Y}}-\ptt{\bar{S}} line in the (110) surface BZ, $k_z=0$, and the Hamiltonian decouples into the eigensectors of the mirror operator $M_z$. For the $+\i $ eigensector, we find
\begin{equation}
H_{(+\i )}=\begin{pmatrix}
M+\tf(k_u^2+k_w^2)&2\tilde{V}(k_u-\i k_w)\\
2\tilde{V}^*(k_u+\i k_w)&-M+\td(k_u^2+k_w^2)
\end{pmatrix}\,,
\end{equation}
where $\tilde{V}=Ve^{-\i \pi/4}$ and a similar block for the $(-\i )$ eigensector:
\begin{equation}
H_{(-\i )}=\begin{pmatrix}
M+\tf(k_u^2+k_w^2)&2\tilde{V}^*(k_u+\i k_w)\\
2\tilde{V}(k_u-\i k_w)&-M+\td(k_u^2+k_w^2)
\end{pmatrix}\,.
\end{equation}
In the following we focus on the $+\i $ sector. To obtain the edge states, we replace $k_w\rightarrow -\i \partial_w$ and use an exponentially decaying ansatz in the $-w$ direction
\begin{equation}
\psi_\lambda(w,k_u)=e^{\lambda(k_u) w}\begin{pmatrix}
\phi_{\lambda}(k_u)\\
\chi_{\lambda}(k_u)\\
\end{pmatrix}\,.
\end{equation}
The secular equation yields four solutions $\beta\lambda_{\alpha}$ with $\beta=\pm1$, $\alpha=1, 2$ and\cite{zhou_finite_2008}
\begin{subequations}
\begin{align}
\label{eq:lambdaF}
\lambda_1^2&=k_u^2+F-\sqrt{F^2-(M^2-E^2)/(t_-^2-t_+^2)}\,,\\
\lambda_2^2&=k_u^2+F+\sqrt{F^2-(M^2-E^2)/(t_-^2-t_+^2)}\,.
\end{align}
\end{subequations}
Here, we have defined
\begin{subequations}
\begin{align}
F&=\frac{Et_+-Mt_-+2V^2}{t_-^2-t_+^2}\,,\\
\tf&=t_+-t_-\,,\\
\td&=t_++t_-\,.
\end{align} 
\end{subequations}
The corresponding spinor wave functions are
\begin{equation}
\psi_{\beta,\alpha}(w)=e^{\beta\lambda_{\alpha} w}
\begin{pmatrix}
1\\
\frac{E-M-\tf(k_u^2-\lambda_{\alpha}^2)}{2\tilde{V}(k_u-\beta\lambda_{\alpha})}
\end{pmatrix}\,.
\end{equation}
Surface states at the $w=0$ surface are exponentially decaying solutions in the $-w$ direction. We therefore make a linear superposition of the $\beta=+1$ states:
\begin{equation}
\Psi(w)=a \psi_{+,1}(w) +b \psi_{+,2}(w)\,.
\end{equation}
We impose the boundary condition
\begin{equation}
\Psi(w=0)=0\,.
\end{equation}
For a nontrivial solution, we have $\lambda_1\neq\lambda_2$, and the energy of the surface states has to satisfy
\begin{subequations}
\begin{align}
M-E&=(-\tf)(\lambda_1-k_u)(\lambda_2-k_u)\,,\\
M+E&=\td(k_u+\lambda_1)(k_u+\lambda_2)\,.
\end{align}
\end{subequations}
If $M>0$, $\td>0$ and $\tf<0$, the solution to these equations is given by
\begin{equation}
E_{(+\i )}(k_u)=M\frac{t_+}{t_-}+ \frac{2|V|\sqrt{t_-^2-t_+^2}}{t_-}k_u\,.
\end{equation}
It follows that the edge state velocity is
\begin{equation}
v=\frac{2|V|\sqrt{t_-^2-t_+^2}}{t_-}=\frac{4|V|\sqrt{-\tf\td}}{\td-\tf}>0\,.
\end{equation}
This is what we wanted to show. In the \ptt{XM}-inverted phase, the mirror Chern number for eigensector $(+\i )$ is $-1$. But because $v>0$, there must be two additional crossings along the line \ptt{\bar S}-\ptt{\bar{Y}}-\ptt{\bar{S}}.

\section{Mean-field treatment of Kotliar-Ruckenstein slave bosons}
\label{sec:kotliar-ruckenstein}
We extend the Hilbert space of our system by introducing slave bosons with annihilation operators $\hat e_i$, $\hat s_{i\sigma}$, and $\hat d_i$ for empty, singly occupied, and doubly occupied $f$ orbitals at site $i$, respectively. The fermionic creation and annihilation operators $f_{i\sigma}^\dagger$ and $f^{}_{i\sigma}$ are replaced by
\begin{subequations}
\begin{equation}
f_{i\sigma}^\dagger\to f_{i\sigma}^\dagger\hat z_{i\sigma}^\dagger\,,\qquad 
f^{}_{i\sigma}\to f^{}_{i\sigma}\hat z_{i\sigma}^{}\,,\label{eq:fermion_hopping}
\end{equation}
where we defined the boson hopping operator
\begin{equation}
\hat z_{i\sigma}:=\hat s_{i\bar{\sigma}}^\dagger \hat d_i^{}+\hat e_i^\dagger \hat s_{i\sigma}^{}\,.
\end{equation}\label{eq:boson_hopping}%
\end{subequations}

The physical subspace of this extended Hilbert space is recovered by imposing the constraints
\begin{subequations}
\begin{gather}
\hat n_i^e+\hat n_{i\uparrow}^s+\hat n_{i\downarrow}^s+\hat n_i^d=1\label{eq:const1}\,,\\
\hat n_{i\sigma}^s+\hat n_i^d=f_{i\sigma}^\dagger f_{i\sigma}^{}\,,\label{eq:const2}
\end{gather}\label{eq:const}%
\end{subequations}%
where we use the number operators $\hat n^a=\hat a^\dagger \hat a$ for $a\in\{e,s,d\}$. The constraint~\eqref{eq:const1} states that any site must be either empty, singly occupied, or doubly occupied. The second constraint~\eqref{eq:const2} connects the presence of an $f$ electron to a singly or doubly occupied site.

As long as the constraints~\eqref{eq:const} are imposed exactly, there exist different choices for the boson hopping operators which produce the same physical spectrum. It has been shown\cite{kotliar_new_1986} that the definition
\begin{align}
\tilde z_{i\sigma}:=\left(1-\hat n_i^d-n_{i\sigma}^s\right)^{-1/2}\,\hat z_{i\sigma}\left(1-n_i^e-n_{i\bar\sigma}^s\right)^{-1/2} \label{eq:boson-hopping}
\end{align}
produces the correct spectrum in the mean-field approximation for the noninteracting case.

In order to simplify the Hamiltonian, we now assume no spatial dependence of the boson operators and in addition replace the them by their expectation values
\begin{subequations}
\begin{alignat}{3}
e&=\langle\hat e_i^{}\rangle=\langle\hat e_i^\dagger\rangle\,,\qquad& n_e&=e^2=\langle\hat n_i^e\rangle\,,\\
s&=\langle\hat s_{i\sigma}^{}\rangle=\langle\hat s_{i\sigma}^\dagger\rangle\,,\qquad& n_s&=s^2=\langle\hat n_{i\sigma}^s\rangle\,,\\
d&=\langle\hat d_i^{}\rangle=\langle\hat d_i^\dagger\rangle\,,\qquad& \nd:=n_d&=d^2=\langle\hat n_i^d\rangle\,.
\end{alignat}\label{eq:mean-field}%
\end{subequations}%
The originally local constraints~\eqref{eq:const} are now only imposed on average and lead to
\begin{subequations}
\begin{align}
n_s&=\nf/2-\nd\,,\\
n_e&=1-\nf+\nd\,.
\end{align}\label{eq:number-operators}%
\end{subequations}%
Note that $\nf$ was defined as $\nf=\langle f_{i\uparrow}^\dagger f_{i\uparrow}^{}\rangle+\langle f_{i\downarrow}^\dagger f_{i\downarrow}^{}\rangle$ and it holds $0\leq\nf\leq2$ while $0\leq n_a\leq 1$ for $a\in\{e,s,d\}$. 

Using~\eqref{eq:mean-field} and \eqref{eq:number-operators}, the boson hopping operators~\eqref{eq:boson-hopping} can now be replaced by
\begin{equation}
z=\frac{\sqrt{\nd\left(\frac{\nf}{2}-\nd\right)}+\sqrt{(1-\nf+\nd)\left(\frac{\nf}{2}-\nd\right)}}{\sqrt{\frac{\nf}{2}\left(1-\frac{\nf}{2}\right)}}\,.
\end{equation}\\
According to \eq{eq:fermion_hopping}, the hopping of $f$ electrons is therefore reduced by a factor $z^2$ and the hybridization by a factor of $z$.
To the resulting mean-field Hamiltonian we have to add a Lagrange multiplier $\lambda$ in order to enforce the relation $\nf=\langle f_i^\dagger f_i^{}\rangle$. This procedure leads to the renormalized noninteracting Hamiltonian described in \Sec{sec:renorm}.

The variables $\lambda$, $\nf$, and $\nd$ can now be determined selfconsistently by the saddle-point equations for the free energy. For $T=0$, these are equivalent to
\begin{equation}
0=\left\langle\frac{\partial\tilde H}{\partial \lambda}\right\rangle=\left\langle\frac{\partial\tilde H}{\partial \nf}\right\rangle=\left\langle\frac{\partial\tilde H}{\partial \nd}\right\rangle\,,
\end{equation}
where the expectation value $\langle\cdot\rangle$ is calculated for the occupied bands. For our model, the three equations can be written as
\allowdisplaybreaks%
\begin{subequations}%
\begin{align}%
0&=1-\nf-S_1\,,\label{eq:spe1}\\
0&=-\lambda+2z\dd{z}{\nf}\,S_2+\dd{z}{\nf}\,S_3\,,\\
0&=U+2z\dd{z}{\nd}\,S_2+\dd{z}{\nd}\,S_3\,,
\end{align}\label{eq:saddle-point-eq}%
\end{subequations}
where $S_1$, $S_2$, and $S_3$ are the expectation values
\begin{subequations}
\begin{align}
S_1&=\frac{1}{N}\sum_k\sum_au_a^\dagger(\bk)\,\frac{\tau_z}{2}\, u_a^{}(\bk)\,,\\
S_2&=\frac{1}{N}\sum_k\sum_au_a^\dagger(\bk)\,\frac{\E-\tau_z}{2}\,h_f(\bk) u_a^{}(\bk)\,,\\
S_3&=\frac{1}{N}\sum_k\sum_au_a^\dagger(\bk)\tau_x \Phi(\bk) u_a^{}(\bk)\,.
\end{align}\label{eq:exp_val}%
\end{subequations}%
Here, the functions $h_f(\bk)$ and $\Phi(\bk)$ are given by \eq{eq:h_d_f_Hyb} and the sums $\sum_a$ run over the two occupied bands.

As the expectation values~\eqref{eq:exp_val} implicitly also depend on $\lambda$, $\nf$, and $\nd$, an analytic treatment of these saddle-point equations is not possible. Instead, we solve them numerically by an iterative method: We solve \eq{eq:spe1} for $\lambda$ as a function of $\nf$ and $\nd$, and use this to minimize the total energy as a function. The expectation values~\eqref{eq:exp_val} are calculated for a discrete set of $\bk$-points.

\pagebreak

\bibliography{Literatur_SmB6}

\end{document}